\documentclass[final]{svjour2}

\smartqed
\usepackage{mathptmx}
\usepackage{times,mathptm}
\usepackage{natbib}
\usepackage{graphicx,epsf,textcomp,pstricks,psfrag}
\usepackage{amsmath}

\begin{document}

%-----------------------------------------------------------------------
% Authors' Macros
%-----------------------------------------------------------------------

%this is "less/similar" symbol
\newcommand{\la}{\mathrel{\mathchoice {\vcenter{\offinterlineskip\halign{\hfil
$\displaystyle##$\hfil\cr<\cr\sim\cr}}}
{\vcenter{\offinterlineskip\halign{\hfil$\textstyle##$\hfil\cr
<\cr\sim\cr}}}
{\vcenter{\offinterlineskip\halign{\hfil$\scriptstyle##$\hfil\cr
<\cr\sim\cr}}}
{\vcenter{\offinterlineskip\halign{\hfil$\scriptscriptstyle##$\hfil\cr
<\cr\sim\cr}}}}}

\newcommand{\be}{\begin{equation}}
\newcommand{\ee}{\end{equation}}
\newcommand{\bea}{\begin{eqnarray}}
\newcommand{\eea}{\end{eqnarray}}

\newcommand{\avk}[1]{{\bf [#1 -- AVK.]}}
\newcommand{\ms}[1]{{\bf [#1 -- M.S.]}}

\newcommand{\rev}[1]{{\bf #1}}

%-----------------------------------------------------------------------
% Title page
%-----------------------------------------------------------------------

\title{Collisional Velocities and Rates\\
       in Resonant Planetesimal Belts}

\titlerunning{Collisions in resonant belts}

\author{
        Martina Queck
        \and
        Alexander V. Krivov      
        \and
        Miodrag Srem\v{c}evi\'c
        \and
        Philippe Th\'ebault
       }

\authorrunning{Queck et al.}

\institute{M. Queck \and A. V. Krivov
             \at Astrophysikalisches Institut, Friedrich-Schiller-Universit\"at Jena,
                 Schillerg\"a{\ss}chen~ 2--3,\\ 07745 Jena, Germany,
                 Tel.: +49 3461 9 47 531
                 Fax:  +49 3461 9 47 532
                 \email{queck@astro.uni-jena.de}
           \and
           M. Srem\v{c}evi\'c 
             \at
             LASP, University of Colorado,
             1234 Innovation Drive, Boulder, CO 80303, USA
           \and
           Ph. Th\'ebault
             \at   
             Stockholm Observatory, Albanova Universitetcentrum, SE-10691 Stockholm, Sweden\\
             and
             LESIA, Observatoire de Paris, F-92195, Meudon Principal Cedex, France
          }

\date{Received May 15, 2007; accepted July 19, 2007}

\maketitle

\begin{abstract}
We consider a belt of small bodies (planetesimals, asteroids, dust particles)
around a star, captured in one of the external or 1:1 mean-motion resonances with a
massive perturber (protoplanet, planet).
The objects in the belt collide with each other.
Combining methods of celestial mechanics and statistical physics, we 
calculate mean collisional velocities and collisional rates, averaged over the belt.
The results are compared to collisional velocities and rates in a similar,
but non-resonant belt,
as predicted by the particle-in-a-box method.
It is found that the effect of the resonant lock on the velocities is rather small,
while on the rates more substantial.
At low to moderate eccentricities and libration amplitudes of tens of degrees,
which are typical of
many astrophysical applications, the collisional rates between objects in an external
resonance are by about a factor
of two higher than those in a similar belt of objects not locked in a resonance.
For Trojans under the same conditions, the collisional rates may be enhanced by up to
an order of magnitude.
The collisional rates increase with the decreasing libration amplitude of the 
resonant argument, depend on the eccentricity distribution of objects, and vary from one 
resonance to another.
Our results imply, in particular, shorter collisional lifetimes of
resonant 
Kuiper belt objects in the solar system and higher efficiency of dust production by 
resonant planetesimals in debris disks around other stars.
\end{abstract}

%-----------------------------------------------------------------------
\section{Introduction}
%-----------------------------------------------------------------------

Resonances between small bodies and giant planets are common in the solar system
and, very likely, in planetary systems around other suns.
Some of the asteroid families in the present-day solar system are locked in various resonances.
Examples are Greeks and Trojans in the 1:1 resonance with Jupiter,
the Hilda group of asteroids in a 3:2 
resonance with it, or the Koronis family in a 5:2 resonance.
In the Edgeworth-Kuiper belt,
Twotinos and Plutinos reside in the 2:1 and 3:2 
resonances with Neptune, respectively.
In the early solar nebula, small planetesimals may have been brought to resonance locations
by gas drag and got  captured in resonances with protoplanets, too
\citep{gold-1975,marzari-weidenschilling-2002}.
Around other stars, resonant belts of planetesimals are thought to be
responsible for clumpy structure observed in resolved debris disks
\citep{wyatt-2003,wyatt-2006,krivov-et-al-2006b}.

Whether resonant or non-resonant, small bodies in all these systems are subject
to collisions, whose typical outcome
varies from one system to another and ranges from perfect agglomeration to
full disruption.
Accordingly, accurate determination of collisional velocities and rates 
amongst the objects is an important ingredient of collisional models.
In non-resonant systems, they can be calculated on the base of the particle-in-a-box approach,
in which the system is followed in a reference frame moving around the star with Keplerian
circular velocity
\citep{oepik-1951,safronov-1969,greenberg-et-al-1978,greenberg-et-al-1991}.
This method has been successfully generalized to include gravitational enhancement,
effects of Keplerian shear, dynamical friction, and viscous stirring
\citep{wetherill-stewart-1989,wetherill-1990}. It has also been implemented as
multiannulus schemes to treat a range of distances around the star 
\citep{spaute-et-al-1991,weidenschilling-et-al-1997}.
Box-based models have been successfully applied for evolution of main-belt asteroids 
\citep{campo-et-al-1994b,bottke-et-al-1994,davis-farinella-1997,bottke-et-al-2005},
Kuiper-belt objects \citep{stern-1995,stern-1996},
debris disks of other main-sequence stars \citep{thebault-et-al-2003},
and accumulation of planetesimals during planetary system formation
\citep[e.g.][and references therein]{lissauer-1993}.

However, the box method essentially involves an assumption that the angular orbital elements of the 
particles (argument of pericenter and mean anomaly) have a uniform distribution.
Accordingly, the population of objects in question has a 
rotationally-symmetric structure.
This basic assumption is violated in the case of a resonance locking.
For particles locked in a resonance, the angles vary in such a way that a certain
combination of them, called resonant argument, librates around a certain value
with a certain amplitude.
As a result, the resonant population may be clumpy.
One might expect the collisional rates, and possibly relative velocities,
of particles in such a clumpy structure to systematically deviate from those 
predicted by the particle-in-a box-method.

The problem can, and has been, studied with direct $N$-body codes.
The difficulty that the ``real'' number of bodies strongly exceeds a
numerically affordable one can be overcome in the following way
\citep[see, e.g.,][and references therein]{thebault-doressoundiram-2003}.
One treats a limited number $N_{num}$ of test particles and assigns to these particles 
an inflated radius such that the total optical depth of the numerical system
is equal that of the ``real'' one: $N_{num} R_{num}^2 = N_{real} R_{real}^2$.
Another possibility \citep{wyatt-2006} is to use a sort of a ``local'' particle-in-a-box
method. An $N$-body code picks up a particle at a location of interest and looks at
all particles in its vicinity. Then their velocities relative to the target particle
and their total cross section are calculated. Finally, a particle-in-a box-like approximation
in that local region gives the collisional rates there.

The aim of this paper is to make an analytic calculation of collisional rates
and collisional velocities within a family of objects locked in one of the mean-motion
resonances with a nearby perturber.
Our general method, which combines classical celestial mechanics with statistical physics,
was published earlier
\citep{krivov-et-al-2005,krivov-et-al-2006, krivov-et-al-2006b}.
The specific 
approach used
here, with a focus on non-uniformly
distributed angles, bears close resemblance to the approach
by Dell'Oro and collaborators 
\citep{delloro-paolicchi-1998,delloro-et-al-1998,delloro-et-al-2001,delloro-et-al-2002}.

However, the latter is rather tuned to explore collisions
in a finite set of 
individual objects with known orbits (e.g., a given asteroidal family), whereas this
paper analyzes ``infinite'' set of fiducial objects with a continuous distribution of orbital
elements.
First, for two arbitrary 
particles in the resonant family, we formulate the collisional condition,
compute the collisional probability per unit time, and
calculate the relative velocity of the two particles at the collision point.
Second, we derive the distributions of the angular elements of the objects from the 
condition of the resonant libration.
Finally, we obtain the average collisional velocities and rates in the whole ensemble
by integrating relevant quantities over all possible pairs of 
colliders, weighted with the ``resonant'' distributions.

Section 2 describes the resonant family of objects:
essential features of the resonant dynamics, simplifying assumptions that we make,
choice of variables etc.
Section 3  deals with a binary collision between two given particles. Here, 
the  collisional condition is derived and the velocity of collision is evaluated.
Section 4 defines, in the form of integral formulas,
collisional velocities and  rates  in the whole 
ensemble of objects. These are calculated in Sections 5 and 6, respectively.
Section 7 addresses the 1:1 resonance case, i.e. Trojans.
Possible applications are discussed in Section 8.
Section 9 contains our conclusions.

%-----------------------------------------------------------------------
\section{Resonant belt}
%-----------------------------------------------------------------------

\subsection{System}

We will consider the following system. There are a massive body, which we will 
call planet, orbiting the primary, referred to as a star, 
and a disk of objects (small bodies or dust),
orbiting the same primary.
Motion of each of the objects or the planet in 3D is described by six  Keplerian 
orbital elements
\begin{eqnarray}
a \; , \; e \; ,\; i \; ,\; \omega \;, \; \Omega \; , \; \lambda,
\end{eqnarray}
which stand for the semimajor axis, eccentricity, inclination, argument of 
pericenter, longitude of ascending node, and mean longitude, respectively.
Instead of $\lambda$, either the mean anomaly $M$ or the true anomaly $\theta$ 
can be used.
The elements of the planet will be marked with a subscript $p$.

\subsection{Simplifications}

To keep the problem analytically tractable, we make several major 
simplifying assumptions.

1. First, as in many  other studies, we assume a {\em circular orbit of the planet}:
$e_p \equiv 0$. Apart from reducing the complexity of the problem dramatically, this assumption
enables comparison with the particle-in-a-box results.
For the solar system, this assumption is natural. For many other systems where planetesimals
are expected to be trapped in resonances as a result of planetary migration, it is reasonable, too,
since dissipative forces that cause migration tend to circularize the planetary orbit
\citep{wyatt-2003}. If dust particles instead of planetesimals are considered, and therefore
the mechanism of resonant capture is dust transport by dissipative forces rather that
planet migration, resonance capture is known to be most efficient for less eccentric planets.
\citet{quillen-2006}, among others,
numerically investigated the dependence of the dust capture probability on
$e_p$. She found that 
resonant capture is only possible if 
$({\cal M}_{\rm p} / {\cal M}_{*} )^{-1/3} e_p \sim 1$,
where $\cal{M}_{\rm p}$ and $\cal{M}_{*}$ are the masses of the planet and the star,
respectively.
Furthermore, even if capture occurs, already a low planetary
eccentricity of $\sim 0.05$ smears a clumpy resonant structure to a
rotationally symmetric ring
%\citep{reche-et-al-2007}.
(Remy Reche, pers. comm).

\begin{figure}
\begin{center}
 \includegraphics[width=8cm]{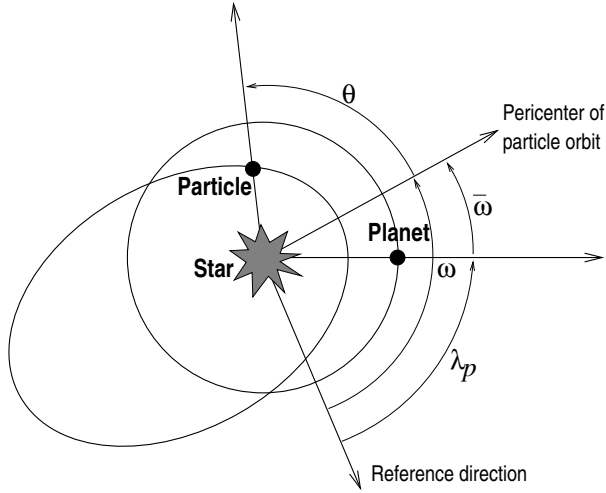} 
 \caption{Angular variables in 2D.
 \label{geometry}
 }
\end{center}
\end{figure}

2. Further, we confine our analysis to {\em small inclinations}:
$i, i_p \la 10^{\circ}$. 
%CM does not know \la
One reason for that is that orbital inclinations of small
bodies in the solar system and other planetary systems have this order of 
magnitude. Another reason is that higher inclinations drastically reduce the
probability of resonance capture or make the resonant orbits unstable
\citep[e.g.,][]{wisdom-1983,jancart-et-al-2003,gallardo-2006}.
Furthermore, we will compute collisional velocities by simply assuming $i = i_p = 0$,
and take into account corrections due to small non-zero
inclinations only in the calculations of collisional rates.
Thus we can assume a 2-dimensional geometry as shown in 
Fig. \ref{geometry} and introduce overlined variables, measured with respect to the planet:
\begin{eqnarray*}
\overline{\omega}  &\equiv& \omega - \lambda_p,\\
%\overline{l}       &\equiv& l - \lambda_p,\\
\overline{\lambda} &\equiv& \lambda - \lambda_p
.
\end{eqnarray*}

3. Next, all particles are assumed to have {\em the same ``typical'' radius}.
This is justified by the fact that we want to focus on geometrical 
effects caused by resonance locking, rather than on the size distribution effects.

In fact, this assumption does not imply any loss of generality, as long as
we consider macroscopic objects, whose dynamics is purely gravitational and therefore 
independent of their size. It would be, however, a simplifying assumption for dust,
the motion of which is affected by radiation pressure and is therefore size-dependent.

4. We consider the resonant population only and thus
assume all objects to have {\em the same, resonant, value of semimajor axis}:
$a \equiv a_{res}$.
Thus the semimajor axis $a$ is treated as a parameter, not a variable.

5. Finally, our analysis is confined to external and the primary resonances only:
$a_{res} \ge a_p$.

%\noindent
Additional, less important simplifications, are introduced below.

\subsection{Mean-motion resonance}

An external mean-motion resonance (MMR) arises when the
periods, being the inverse of mean motions, of a planet $P_p$ and a particle $P$ are in a rational commensurability:
$P/P_p = (p+q)/p$, where $p$ and $q$ are integers.
The MMRs are located at
$a_{res} = a_{p} [ (p+q) / p ]^{2/3}$.

If ``particles'' are small dust grains rather than planetesimals,
$a_{res}$ is shifted by a factor of $(1-\beta)^{1/3}$, where
$\beta$ is the radiation pressure to gravity ratio.
In this paper, we set $\beta$ to zero.

For an object residing in the resonance, a certain combination of 
its orbital elements, called the resonant argument,
\begin{equation*}
\nonumber
\Phi
  = 
  (p+q)\lambda -p\lambda_p - q\omega 
  =
  (p+q)\overline{\lambda} - q\overline{\omega}
\end{equation*}
librates around a certain value $\Phi_0$,
often close to $0^\circ$, $180^\circ$, or $270^\circ$,
with a given amplitude $A$
\citep{murray-dermott-1999,kuchner-holman-2003,wyatt-2006}.
In the case of the primary 1:1-resonance ($p=1$, $q=0$), the
resonant argument simplifies to
$$
\Phi = \overline{\lambda}
$$
and librates around the $\Phi_0 = \pm \pi/3$, corresponding to the motion
in the vicinity of the Lagrange points $L_4$ and $L_5$.

Next, the objects captured in a resonance tend to gradually pump  their 
orbital eccentricities, up to a certain value $e_{max}$. Both the libration
amplitude $A$ and the maximum eccentricity $e_{max}$ depend, in a complex way,
on the mechanism of resonant trapping (Poynting-Robertson effect, planet migration, etc.)
Furthermore, even for a given trapping mechanism, both quantities depend on
a multitude of physical parameters: planet mass, order of the resonance,
etc. Throughout this paper, we assume that both $A$ and $e_{max}$ can be
``pre-determined'' by a dedicated dynamical study of a system of interest,
and therefore treat them as {\em free parameters}.

\subsection{Distributions}

In 2D, four orbital elements fully describe the 
particle's motion: $a$, $e$, $\theta$, $\overline{\omega}$.
As $a$ is a parameter, only three remaining elements
represent the phase space variables.

We now introduce the distributions of our variables.
Following \cite{krivov-et-al-2005}, we denote by
$\phi(x,y,...) dx dy ...$ the fraction of particles in the disk with arguments
$[x, x+dx]$, $[y, y+dy]$, $...$ . The $\phi$-functions are
normalized to unity:
$$
 \int \ldots \int\phi(x,y,...) dx dy ...  = 1.
$$

The mean longitude $\overline{\lambda}$ is distributed uniformly, and the distribution
of the true anomaly  $\theta$ follows from the Kepler equation:
\begin{eqnarray*}
  \phi(\overline{\omega}, \theta ) 
    &=& 
   \phi(\overline{\omega},\overline{\lambda} ) 
   \left|\frac{\partial \overline{\lambda}}{\partial\theta}\right|_{\overline{\omega}}
    =
    \phi(\overline{\omega}, \overline{\lambda})\; 
    \frac{r^2}{a^2 \sqrt{1-e^2}}
\label{philambda}
\mbox{ ,}
\end{eqnarray*}
where $\overline{\lambda}$ in the right-hand side should be calculated from $\theta$
by means of standard formulas of Keplerian motion
\begin{eqnarray}
  \overline{\lambda}
  &=&
  \overline{\omega} + M,
\label{lanbda_M}\\
  M
  &=&
  E - e \sin E,
\label{M_E}\\
  \tan {E \over 2}
  &=&
  \sqrt{1-e \over 1+e}
  \tan {\theta \over 2}
\label{E_theta} .
\end{eqnarray}

Within the resonance, where
$\Phi_0 - A < \Phi < \Phi_0 + A$,
the distributions of $\overline{\omega}$
and $\lambda$ (or $\theta$)
are not independent.
Assuming the distribution of $\Phi$ within the libration width
to be uniform,
we obtain
\begin{equation}
 \phi_{\omega}(\overline{\omega},\overline{\lambda})
 =
 \frac{1}{2\pi}\;
 \frac{1}{2A} \;
  H\left[\Phi_0-A < \Phi <\Phi_0 +A\right] ,
\label{phiom}
\end{equation}
where $H[cond]$ is a Heaviside function, which equals one if the
 evaluation of $cond$
returns {\it true} and zero if $cond$ returns {\it false}.
If needed, $H$ can be replaced by  a more realistic distribution, e.g. sinusoidal.

Particles caught in a resonance may have eccentricities between zero
and a maximum value $e_{max}$.
To keep the analysis simple,
we assume a uniform distribution between the two borders,
\begin{eqnarray}
\phi_e(e; e_{max}) = {1 \over e_{max}} H[e < e_{max}] .
\label{phie}
\end{eqnarray}
This simplification, like the one for the resonant argument made above,
can always be lifted by replacing the Heaviside function
with a more realistic distribution.

\begin{figure*}
\centerline{ \includegraphics[width=0.95\textwidth]{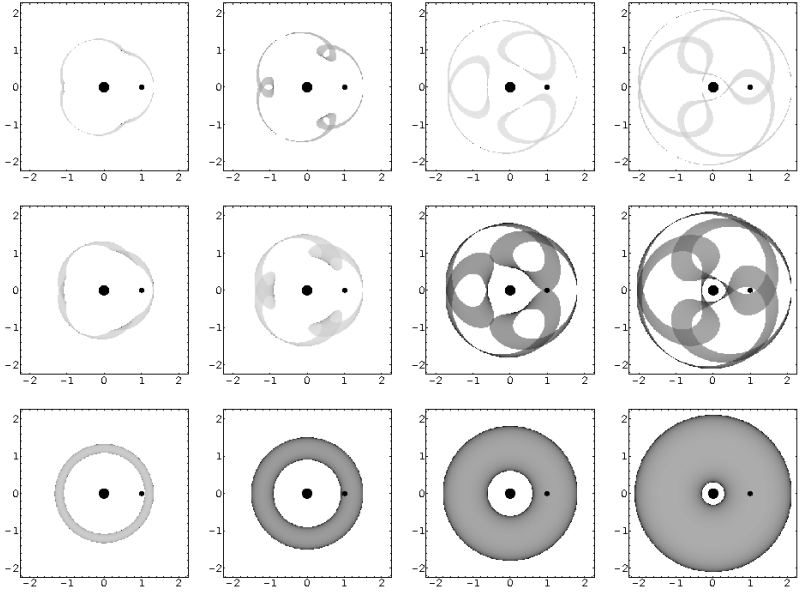}}
\caption{Spatial distribution
of objects, locked in a 4:3 resonance.
The grey scale is arbitrary; the darker the grey, the higher the ``density'' of objects. 
The star at (0,0) and the planet at (1,0) are shown with large and small circles.
From left to right: dependence on eccentricity, $e=0.1$, $0.25$, $0.5$, and $0.75$.
From top to bottom: dependence on the libration amplitude,
$A=0.1\pi$ ($18^\circ$), $0.3\pi$ ($54^\circ$), and $\pi$ ($180^\circ$).
 \label{garden}
 }
\end{figure*}

Figure~\ref{garden} gives a visual impression of how the distribution (\ref{phiom}) describes
a resonance population.
Plotted is the distribution of cartesian coordinates
$\phi_{xy}(x,y)$ calculated from
$\phi_\omega(\overline{\omega},\overline{\lambda})$
by the virtue of
\be
\phi_{xy}(x,y) |dx dy| = 
\phi_\omega(\overline{\omega},\overline{\lambda})
|d\overline{\omega} \, d\overline{\lambda}| .
\ee

\begin{figure*}
\centerline{ \includegraphics[width=\textwidth]{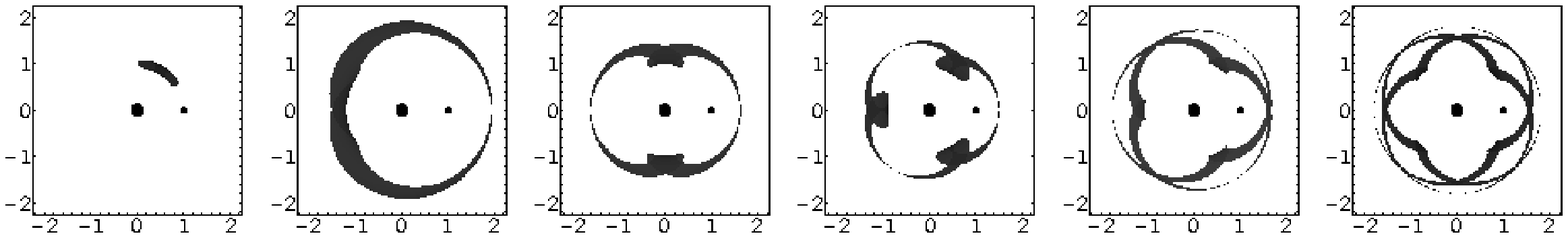}}
\caption{
Same as Fig. \ref{garden}, but for objects with
the same orbital eccentricity
of $e=0.25$ and libration width of $A=0.2\pi$ ($36^\circ$), locked in 
different resonances.
From left to right: 1:1 (Trojans near $L_4$), 2:1, 3:2, 4:3, 5:3, 7:4.
 \label{garden1}
 }
\end{figure*}

The upper plots, drawn for a small libration width $A$, show
``loopy'', pretzel-like structures, well-known to be typical of the synodic motion of 
resonant particles \citep[cf.][]{murray-dermott-1999,kuchner-holman-2003}.
The middle panels show how a larger libration amplitude dithers the distribution.
Yet more fuzziness will obviously come after convolving Eq.~(\ref{phiom}) with a distribution of 
eccentricities (\ref{phie}).
Finally, the lowest panels with $A=180^\circ$ that correspond to a non-resonant case become
rotationally-symmetric. 
Note that the distributions are not radially uniform even in this case:
the particle density is higher at the inner and outer edges of each ring. Mathematically, the density
there gets infinitely large, because the radial velocity of particles vanishes in pericenters and
apocenters.
Figure~\ref{garden1} shows configurations of particles locked in different resonances, from 
\mbox{0-th} to
the 3rd order. For simplicity, the eccentricity and the libration width are the same on all panels:
$e=0.25$ and $A=0.2\pi$ ($36^\circ$).

%-----------------------------------------------------------------------
\section{Collisions}
%-----------------------------------------------------------------------

\subsection{Collision condition}

In terms of radius vectors, the collision  condition
is trivial: $\mathbf{r}_1 = \mathbf{r}_2$,
meaning that distances and true longitudes $\overline{\omega}+\theta$ 
 of both particles coincide:
\begin{eqnarray}
r_1 &=& r_2 \\
\overline{\omega}_1 + \theta_1 &=& \overline{\omega}_2 + \theta_2 . 
\label{o1t1=o2t2}
\end{eqnarray}
By applying the equation of conic section and solving for
$\overline{\omega}_2$ and $\theta_2$,
%calling the solutions
%$\overline{\omega}_2^c$ and $\theta_2^c$. They are
we get
\begin{eqnarray}
\overline{\omega}_2^{c} &=& \theta_1 - \theta_2^{c} + \overline{\omega}_1
\label{omega_2^c}
\\
\cos\theta_2^c            &=& {1\over e_2} 
                          \left[
			  {a_2 \over a_1}
			  {1-e_2^2 \over 1-e_1^2}
			   \left(
			   1+e_1 \cos\theta_1
			   \right)
			   -1
                          \right],
\label{theta_2^c}
\end{eqnarray}
which splits up into two solutions (``$+$'' and ``$-$'')
\begin{eqnarray}
  \overline{\omega}_2^{+} = \theta_1 - \theta_2^{+} + \overline{\omega}_1,
  \qquad
  \theta_2^+              &=& \arccos\left[ \cos\theta_2^c \right]
  \label{cc1}
\\
\mbox{ and}\hspace*{3.5cm}&& \nonumber\\
  \overline{\omega}_2^{-} = \theta_1 - \theta_2^{-} + \overline{\omega}_1 ,
  \qquad
  \theta_2^-              &=& 2\pi - \arccos\left[ \cos\theta_2^c \right] .
  \label{cc4}
\end{eqnarray}

\subsection{Relative velocity at collision}

To compute the relative velocity 
$V_{imp}^k
 (e_1,\overline{\omega}_1,\theta_1    \;,\;   e_2,\overline{\omega}_2,\theta_2)$ 
of two colliding particles,
we start with calculating the velocity vector of either particle. 
Consider a cartesian coordinate system centered on the star and with
$x$-axis pointing towards the particle's present position,
but \emph{not} rotating with the radius vector.
The particle's position and velocity in this system are
\begin{eqnarray}
\textbf{r}   &=& \left( \begin{tabular}{c}
                     $r$ \\ 0 
                    \end{tabular} \right) , \\
\textbf{v}   &=&  \left( \begin{tabular}{c}
                     $v_r$ \\ $v_{\phi}$ 
                    \end{tabular} \right) 
	  = \sqrt{GM \over a(1-e^2)} \left( \begin{tabular}{c}
                     $e\sin\theta$ \\ $1+e\cos\theta$ 
                    \end{tabular} \right) . 
\label{v}
\end{eqnarray}
Denoting the velocity vectors of two colliders
by $\textbf{v}_1$ and $\textbf{v}_2$, the $k$-th power of the relative velocity is
\begin{equation}
V_{imp}^k = |\textbf{v}_1 - \textbf{v}_2|^k 
.
\label{vimp^k}
\end{equation}
Therefore, the relative velocity can be calculated by simply
applying Eq.~(\ref{v}) to both colliders, which makes it obvious that
$V_{imp}=V_{imp}(e_1,\theta_1,e_2,\theta_2)$.
Furthermore, at collision, $\theta_2$ is determined by 
$e_1$, $\theta_1$, and $e_2$, see Eq.~(\ref{theta_2^c}).
Therefore, the impact velocity depends on
three arguments only:
\begin{equation}
  V_{imp}=V_{imp}(e_1,\theta_1,e_2).
 \label{vimp^1}
\end{equation}
We do not give the explicit form of $V_{imp}^k$,
because it is rather lengthy and obtained by a straightforward  calculation.

%-----------------------------------------------------------------------
\section{General formalism}
%-----------------------------------------------------------------------

\subsection{Splitting of variables}

Following \citet{krivov-et-al-2005,krivov-et-al-2006},
we now arrange all phase space variables into ``useful'' ones $\mathbf{p}$
that we keep and ``dummy'' ones
$\mathbf{q}$ that  we will average over.
For our problem, we choose
\begin{eqnarray}
{\bf p} = (e) 
\label{p_def},
\qquad
{\bf q} = (\overline{\omega},\theta) .
\label{q_def}
\end{eqnarray}
Recall that $a$ is a parameter, not a phase space variable.

\subsection{$\Delta$-integrals}

Consider an arbitrary function defined at the collisional point,
for instance the $k$-th power of the relative velocity of two colliding
particles $V_{imp}^k (\textbf{p}_1,\textbf{q}_1,\textbf{p}_2,\textbf{q}_2)$.
\citet{krivov-et-al-2005,krivov-et-al-2006} have shown that the mean
value of that function, averaged over all ${\bf q}$-variables, can be
expressed as
\begin{eqnarray}
V_{imp}^k(\textbf{p}_1,\textbf{p}_2)
&=&
{
 \Delta^{(k)}(\textbf{p}_1,\textbf{p}_2)
   \over
 \Delta^{(0)}(\textbf{p}_1,\textbf{p}_2)
} .
\label{vimp_e1e2_general}
\end{eqnarray}
where 
\begin{equation}
\begin{split}
 &\Delta^{(k)}(\textbf{p}_1,\textbf{p}_2)
 \equiv\\
& \int_{\textbf{q}_1} \int_{\textbf{q}_2}
 V_{imp}^k\left(\textbf{p}_1,\textbf{q}_1,\textbf{p}_2,\textbf{q}_2 \right)
 \delta\left(\textbf{r}_1 - \textbf{r}_2 \right)
 \phi_q(\textbf{q}_1)\phi_q(\textbf{q}_2) d\textbf{q}_1 d\textbf{q}_2 
.
\end{split}
\label{delta_def}
\end{equation}
This formula already includes the collisional condition
through the factor 
\mbox{$\delta (\textbf{r}_1 - \textbf{r}_2 )$}.
We will refer to $\Delta^{(k)}$ as $\Delta$-integral.

With our choice of variables and setting $k=1$, Eq.~(\ref{vimp_e1e2_general}) 
can be rewritten as
\begin{eqnarray}
V_{imp}(e_1,e_2) &\equiv& {\Delta^{(1)}(e_1,e_2) \over \Delta^{(0)}(e_1,e_2)} ,
\label{vimp_e1e2}
\end{eqnarray}
which gives the average collisional velocity between two overlapping rings of particles:
one with eccentricity $e_1$ and another one with eccentricity $e_2$, with angular
elements of particles distributed in accordance with the resonance condition.

\subsection{Meaning of $\Delta$-integrals}

As was shown in \cite{krivov-et-al-2005}, $\Delta^{(0)}$ can be interpreted as
the reciprocal of  an ``effective interaction volume''.
Consider again two rings formed
by two subsets of particles with given eccentricities $e_1$ and $e_2$.
If $S_1$ and $S_2$ are the surface areas of the rings $e_1$ and $e_2$, and
$S_{12}$ the area of their intersection, then
the zeroth integral $\Delta^{(0)}(e_1,e_2)$ is approximately
\[
\Delta^{(0)}(e_1,e_2) \approx { S_{12} \over S_1 S_2}.
\]

The integral $\Delta^{(1)}$ allows direct physical interpretation, too.
Since 
$\Delta^{(1)} = V_{imp} \Delta^{(0)}$ 
$\sim {\rm  velocity} / {\rm volume}$,
one expects that, after multiplication by the number of particles and their
collisional cross section, $\Delta^{(1)}$ would give the collisional rate.
More precisely, this will be the rate of collisions between particles
with eccentricity $e_1$ and those with eccentricity $e_2$.
This is indeed the case, see Eqs.~(\ref{R2Dinitial})--(\ref{R2D}) below.

\subsection{Evaluation of $\Delta$-integrals}

For the actual calculation of any $\Delta$-integral we insert \eqref{vimp^1} and
\eqref{p_def} into \eqref{delta_def}:
\begin{eqnarray}
 \Delta^{(k)}(e_1,e_2)
 &=&
  \int_{\overline{\omega}_1} \int_{\overline{\omega}_2}
  \int_{\theta_1} \int_{\theta_2}  
 V_{imp}^k (e_1, \theta_1, e_2 )
\nonumber\\
&\times&
 \delta\left(\textbf{r}_1 - \textbf{r}_2 \right)
%\nonumber
\label{delta_var}
\\
\nonumber
&\times&
 \phi_{\overline{\omega}}(\overline{\omega}_1,\theta_1)\;
 \phi_{\overline{\omega}}(\overline{\omega}_2,\theta_2)\;
 d\overline{\omega}_1 d\overline{\omega}_2 \;
 d\theta_1 d\theta_2
.
\end{eqnarray}
The $\delta$-function in \eqref{delta_var} should now be expressed through
orbital elements:
\begin{equation}
\delta\left(\textbf{r}_1 -\textbf{r}_2\right)=
       J^{-1}\delta \left(\overline{\omega}_1 -\overline{\omega}_2 \right) 
             \delta \left(\theta_1 -\theta_2\right) 
\mbox{ ,}
\label{thisisdiracdelta}
\end{equation}
where the Jacobian 
\[
J= \left|
     \frac {\partial\textbf{r}_2} {\partial\left(\overline{\omega}_2,\theta_2\right)}
   \right|
%\mbox{ .}
\]
explicitly reads
\begin{equation}
J = {r^3 \over a} { e_2 \over 1-e_2^2 } |\sin \theta_2|
  = a^2 {e_2 (1-e_2^2)^2 |\sin \theta_2| \over (1+e_2 \cos\theta_2)^3}
\mbox{ .}
\end{equation}
Collecting all intermediate results and incorporating them into \eqref{delta_var}, we get
\begin{eqnarray}
 \Delta^{(k)}(e_1,e_2)
 &=&
  \int_{\overline{\omega}_1} \int_{\overline{\omega}_2}
  \int_{\theta_1} \int_{\theta_2}
 V_{imp}^k (e_1, \theta_1, e_2 )
\nonumber\\
&\times&
{1 \over a^2}
{(1+e_2 \cos\theta_2)^3 \over e_2 (1-e_2^2)^2 |\sin\theta_2| }
\delta\left(\overline{\omega}_2-\overline{\omega}_2^c\right)
\delta(\theta_2-\theta_2^c)
\nonumber\\
&\times&
 \phi_{\overline{\omega}}(\overline{\omega}_1,\theta_1)\;
 \phi_{\overline{\omega}}(\overline{\omega}_2,\theta_2)\;
 d\overline{\omega}_1 d\overline{\omega}_2 \;
 d\theta_1 d\theta_2
.
\label{deltak_general}
\end{eqnarray}

In Appendix A, this formula for the $\Delta$-integral is transformed further.
Three of the four integrals turn out to be analytically solvable, so that
the final expression, Eq.~(\ref{deltak_final}), contains only an integral
over $\theta_1$, which we evaluated numerically.

The $\Delta$-integrals, and therefore the collisional velocities and rates,
have several useful properties, whose derivation
is given in the Appendices A and B:

1. The $\Delta$-integral is symmetric with respect to its arguments:
$\Delta^{(k)}(e_2,e_1)=\Delta^{(k)}(e_1,e_2)$.

2. The  $\Delta$-integral depends on the libration width $A$,
but does not depend on the libration center $\Phi_0$.

3. The limit of the $\Delta$-integral at $A \rightarrow 0$ is finite, and close
to the values obtained with small, but non-zero $A$.

4. The same holds for the limit of the $\Delta$-integral at $e_1 \rightarrow e_2$:
it is finite and is not far from the values calculated for close, but not equal
$e_1$ and $e_2$.

\subsection{Comparison with approach by Dell'Oro et al. }

The formalism developed here is based on exactly the same ideas as the one
by Dell'Oro and collaborators \citep{delloro-paolicchi-1998,delloro-et-al-1998}.
For instance, our $\Delta^{(1)}$-integral (\ref{delta_def}) is essentially the same
as Eq.~(9) or (10) in \citet{delloro-paolicchi-1998}.
A technical difference between the two approaches is that we incorporate the collisional
condition directly into the integrand, through the function $\delta (\textbf{r}_1 - \textbf{r}_2 )$,
and assume a particular functional form of the distribution of orbital elements
(Eqs.~\ref{phiom} and \ref{phie}).
We then perform analytically as many integrations in the multiple integral
as possible. As a result, only one integration (see Eq. \ref{deltak_final})
needs to be performed numerically.
The price to pay for ``more analytics''
in our approach is that it is much ``heavier'' mathematically, which makes a strict
3D-treatment impossible.
Still, we deem this approach suitable for theoretical study of a statistical
ensemble of pseudo-objects with continuous distributions of orbital elements.
As we will see, our approach is quite convenient to explore dependence of the collisional
velocities and rates on various parameters ($e_{max}$, $A$, $p$, and $q$). Besides,
our approach naturally circumvents numerical difficulties that otherwise
would arise for extreme values of these parameters (e.g., for very low or very high
eccentricities).

In contrast, Dell'Oro et al. use a Monte-Carlo technique to evaluate the multiple
integrals, similar to our $\Delta^{k}$.
This allows a study in three dimensions; in fact, they use
\begin{eqnarray}
{\bf p} = (a, e, i)
\qquad
{\bf q} = (\Omega, \omega, \theta)
\end{eqnarray}
(cf. Eq. \ref{q_def}).
This makes their approach ideal for study of particular collisional complexes
in the solar system, consisting of individual objects with known orbital elements.
Their method is particularly useful to explore effects associated
with inclinations and longitudes of nodes.

%-----------------------------------------------------------------------
\section{Collisional velocities}
%-----------------------------------------------------------------------

\subsection{Collisional velocity for the subsets of particles with $e=e_1$ and $e=e_2$}

\begin{figure*}[t!]
\psfrag{Vee}{\hspace*{-3ex} $V(e_1,e_2=0.1)$}
\psfrag{e1}{$e_1$}
\begin{tabular}{cc}
 \includegraphics[totalheight=0.25\textheight]{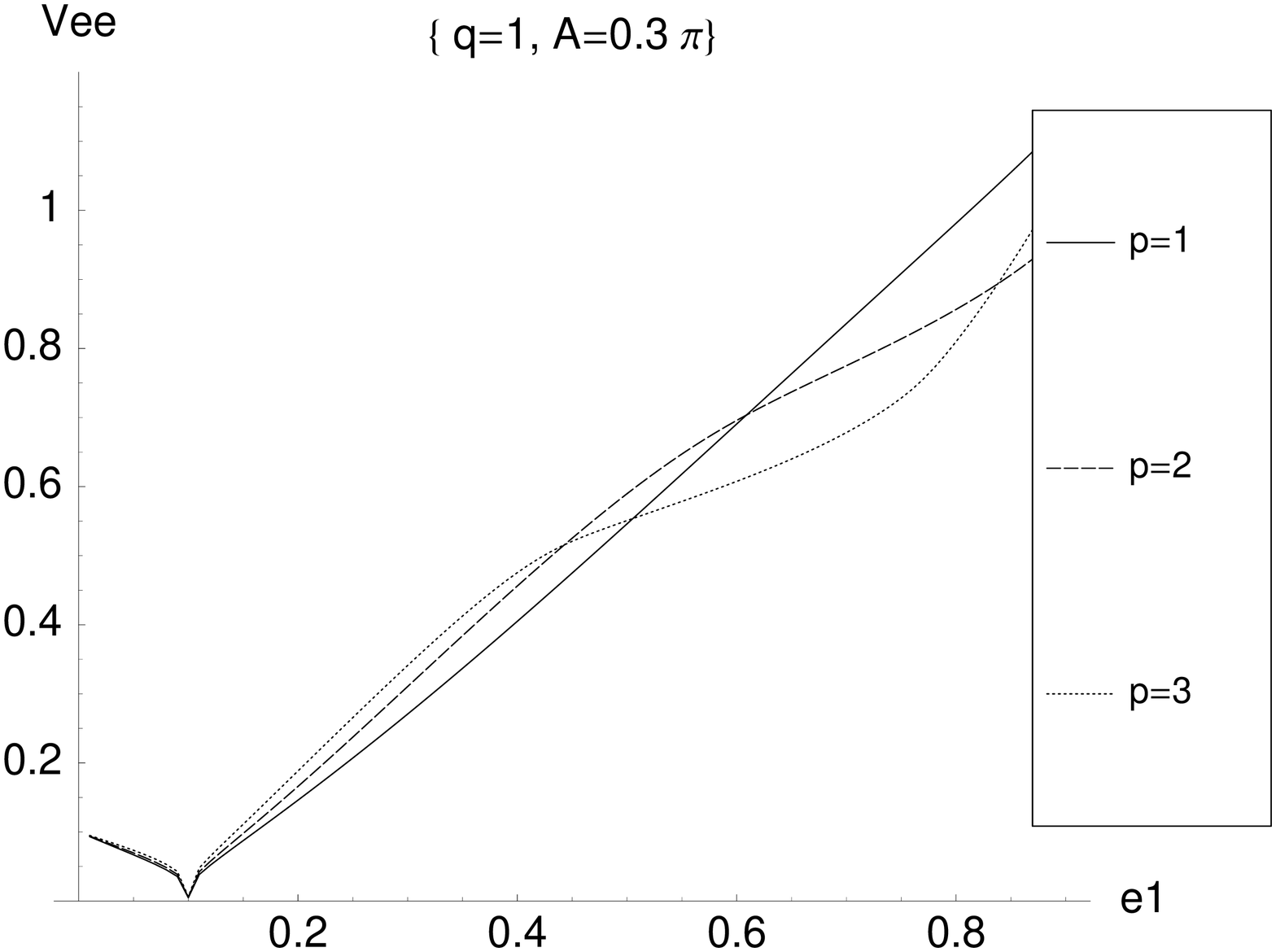} 
 & \\
 \includegraphics[totalheight=0.25\textheight]{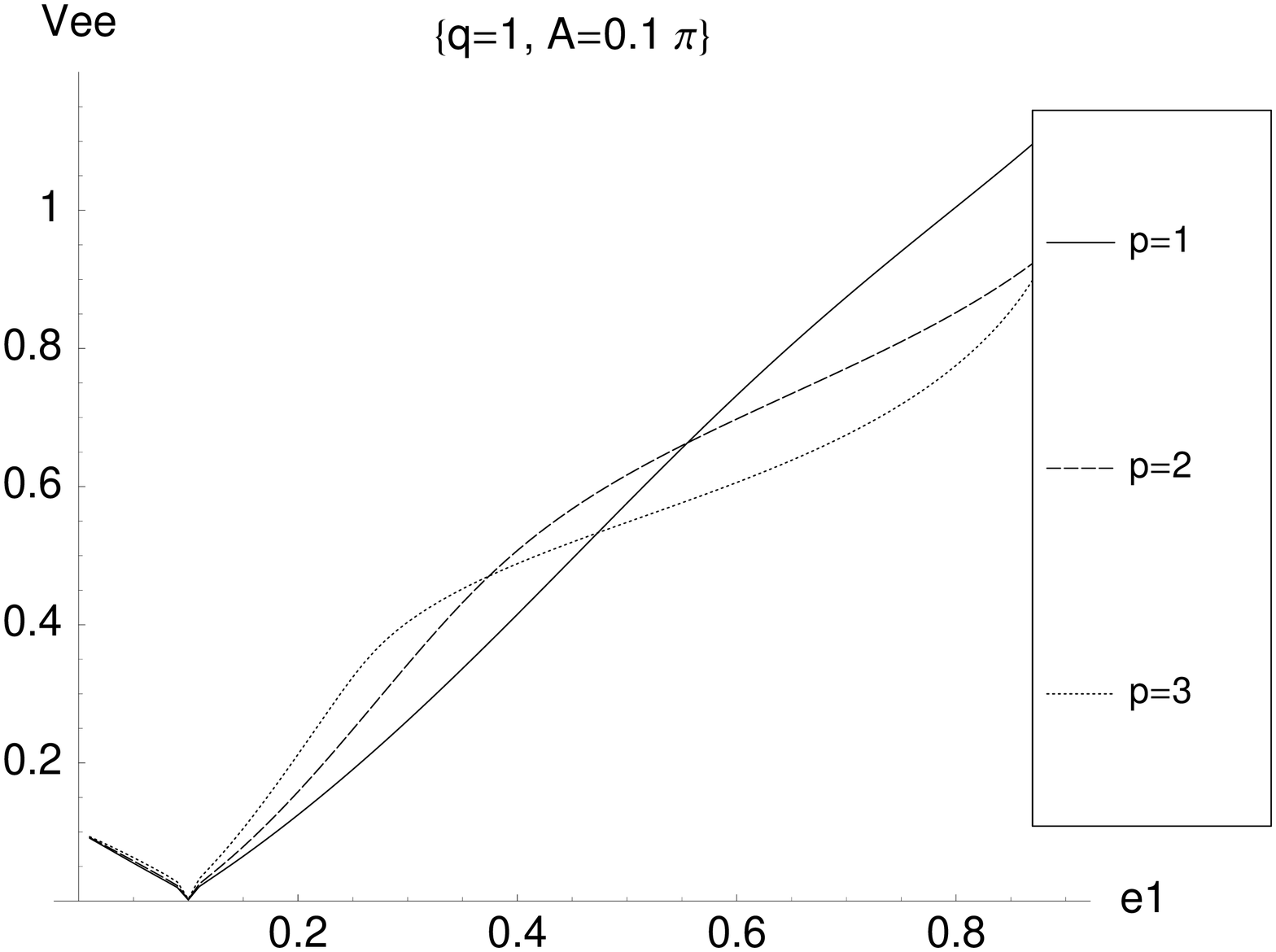}
 &
 \includegraphics[totalheight=0.17\textheight]{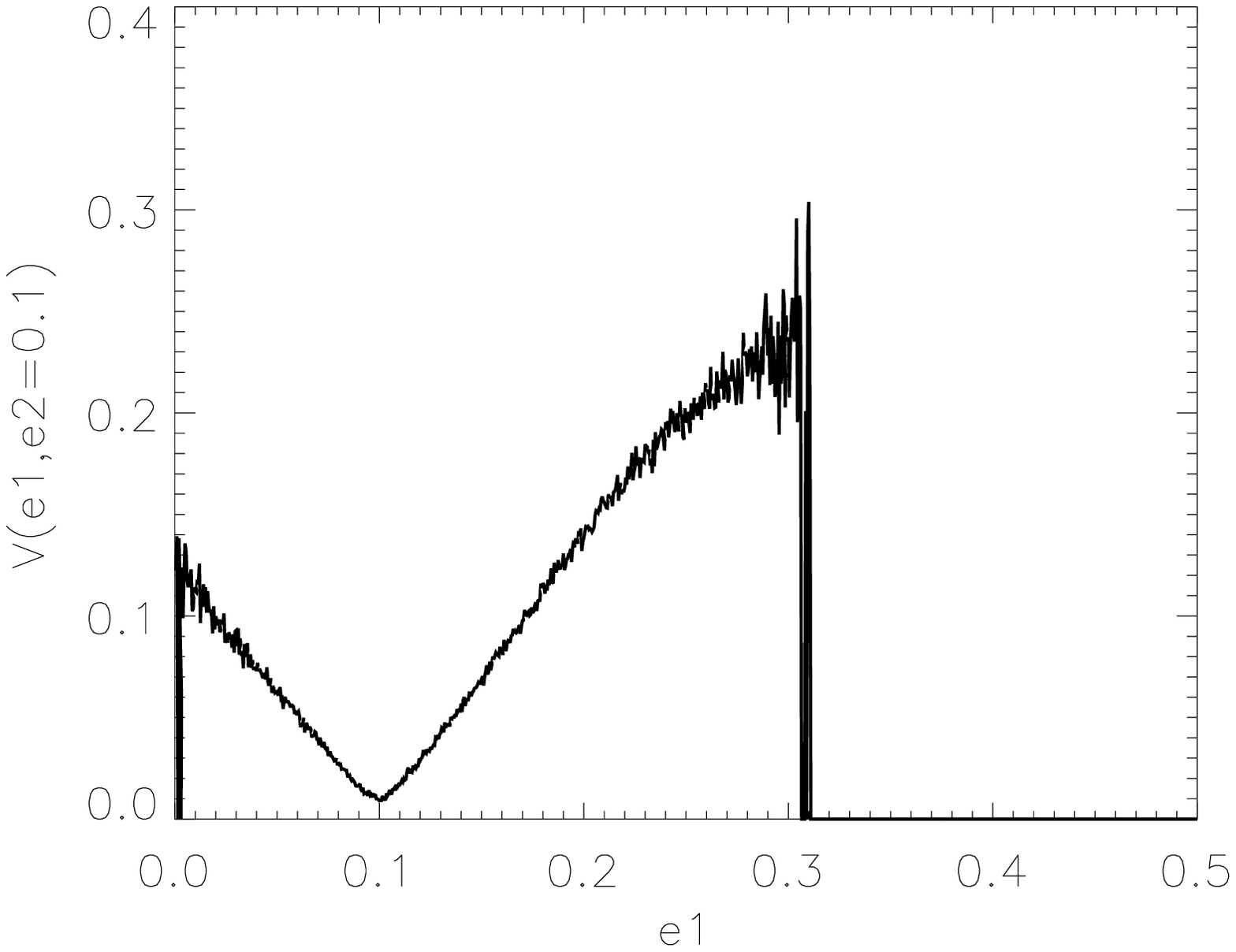}
\end{tabular}
 \caption{ 
 Cuts through the contour plots of collisional velocity $V_{imp}(e_1,e_2=0.1)$,
 in units of $V_{kepler}$,
 for the 2:1 resonance (solid), 3:2 (dashed), 4:3 (dotted).
 Top left: a shallower resonance with $A=0.3\pi$, 
 bottom left: a strong one with $A=0.1\pi$.
 Right panel: typical numerical results
 for $V_{imp}(e_1,e_2=0.1)$ and a 2:1 resonance for comparison.
 \label{fig_vele1}
 }  
\end{figure*}

\begin{figure*}[htb!]
\begin{center}
\psfrag{e2}{$e_2$}
\psfrag{top}{}
\psfrag{bottom}{}
 \includegraphics[totalheight=3.5cm]{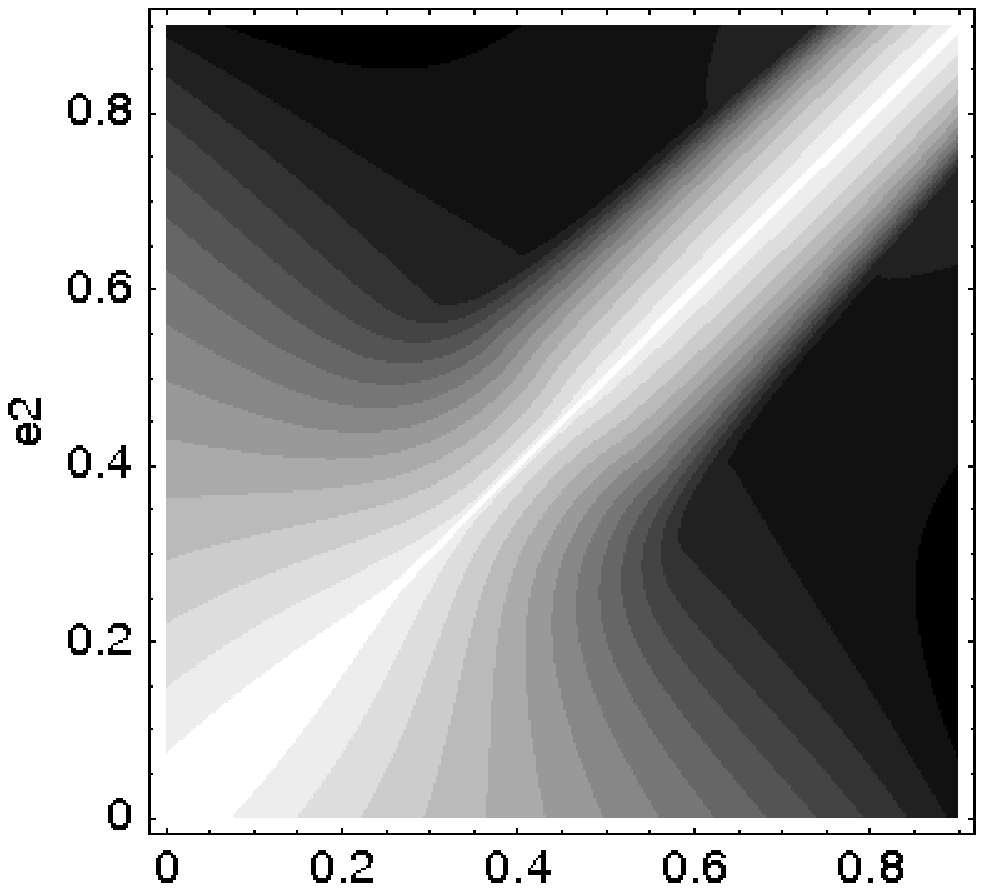}
 \includegraphics[totalheight=3.5cm]{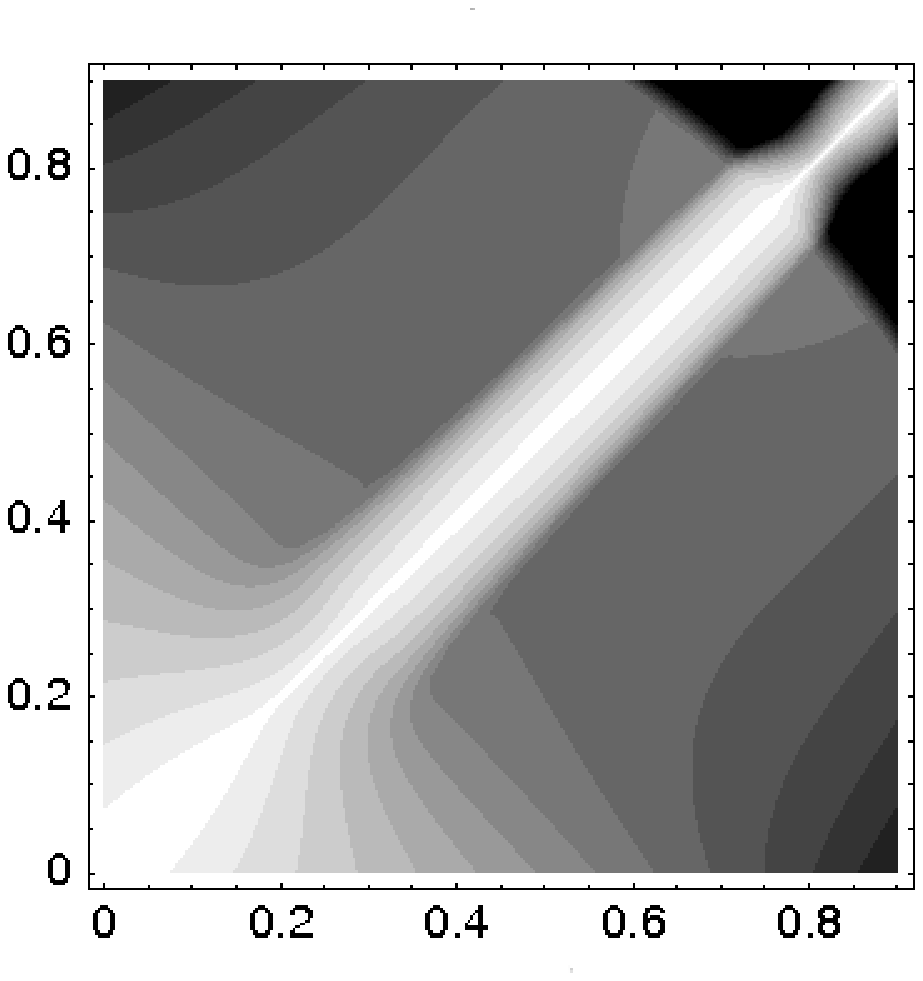}
 \hspace*{-3mm}
 \includegraphics[width=4.5cm,totalheight=3.66cm]{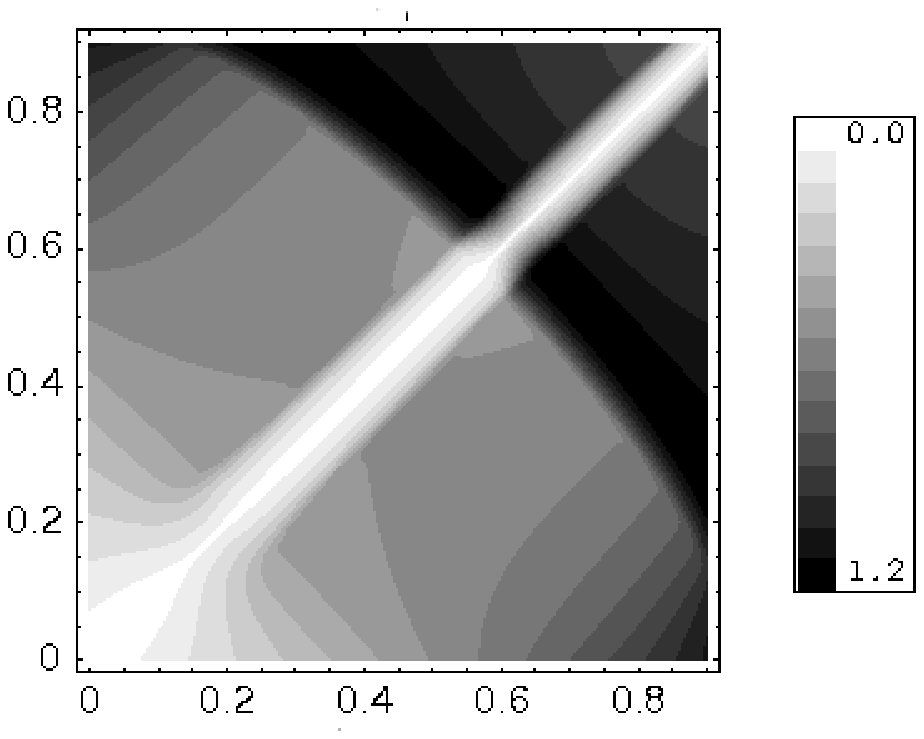}\\[-3mm]
\psfrag{top}{}
\psfrag{bottom}{}
 \includegraphics[totalheight=3.5cm]{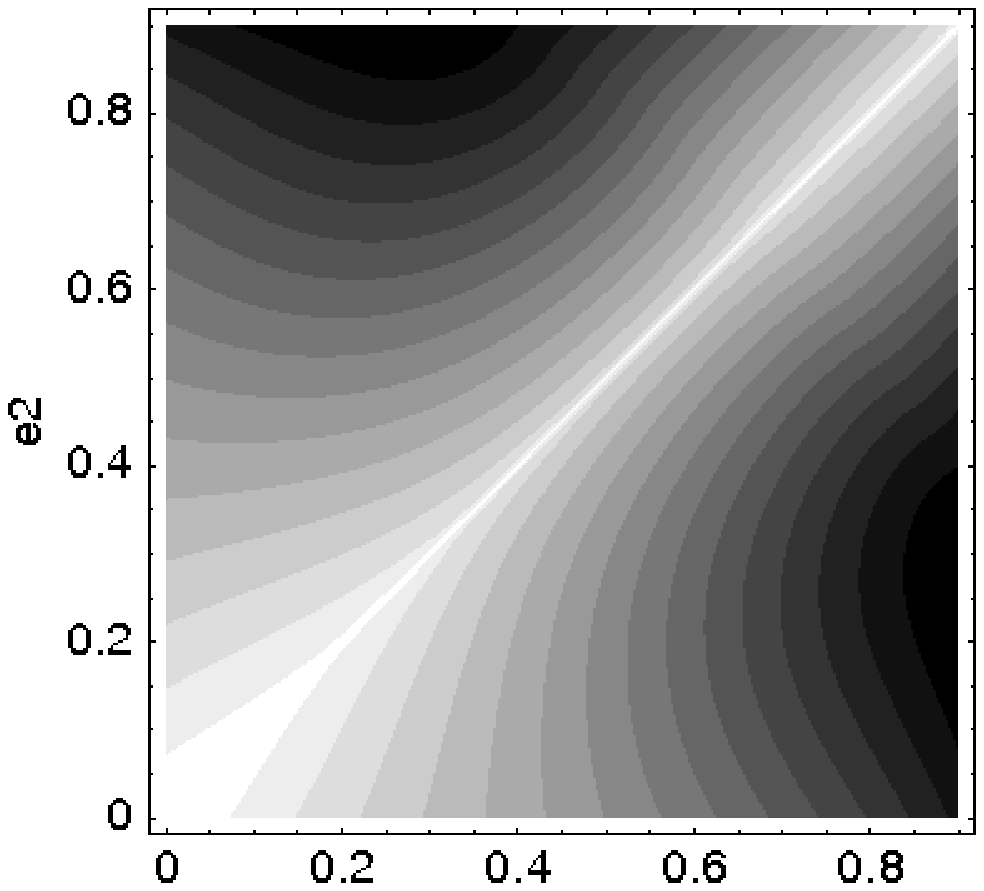}
 \includegraphics[totalheight=3.5cm]{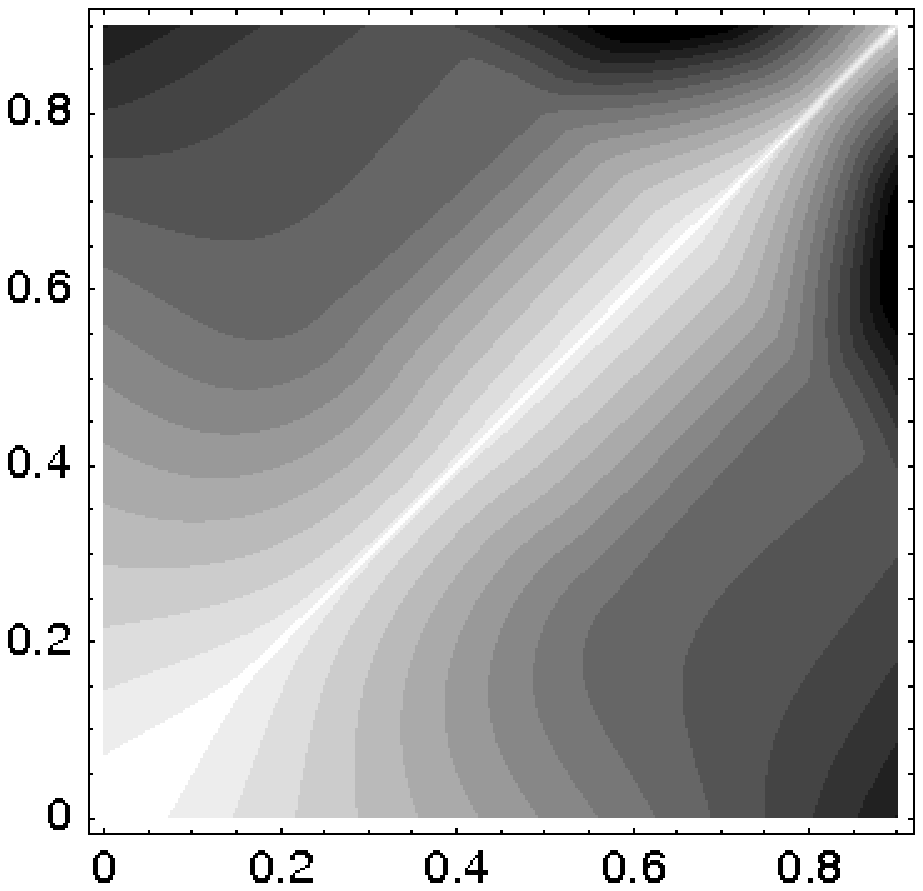}
 \hspace*{-3mm}
 \includegraphics[width=4.5cm,totalheight=3.66cm]{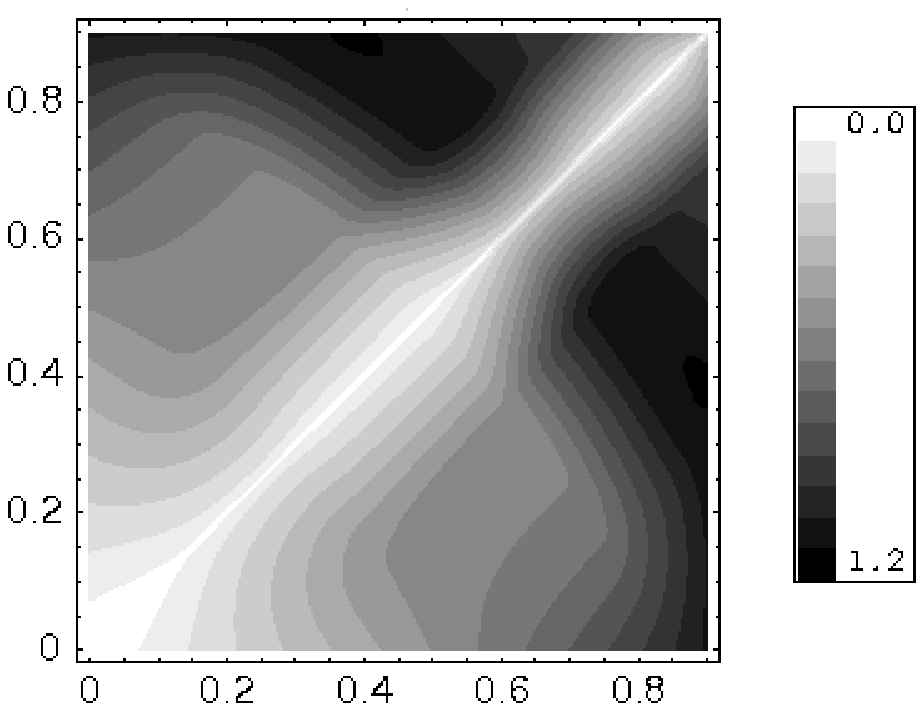}\\[-3mm]
\psfrag{top}{}
\psfrag{bottom}{\hspace*{1.3ex} $e_1$}
 \includegraphics[totalheight=3.5cm]{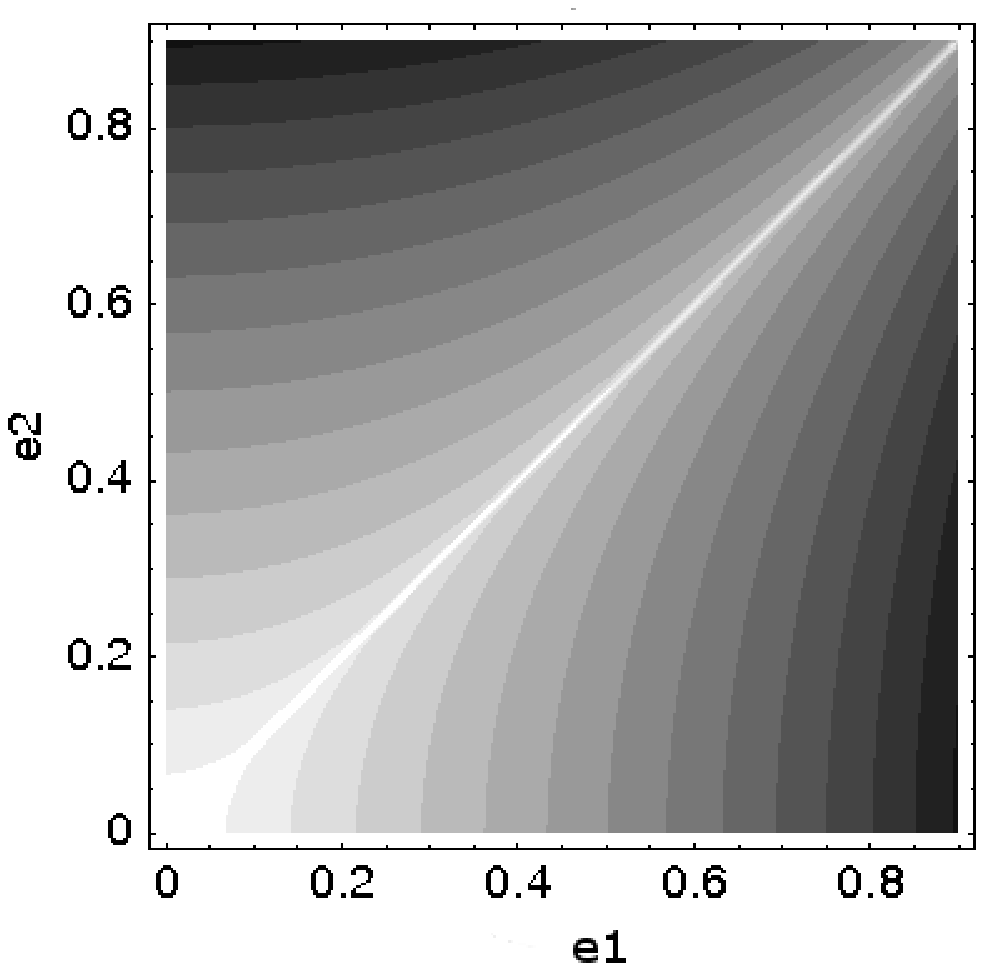}
 \includegraphics[totalheight=3.5cm]{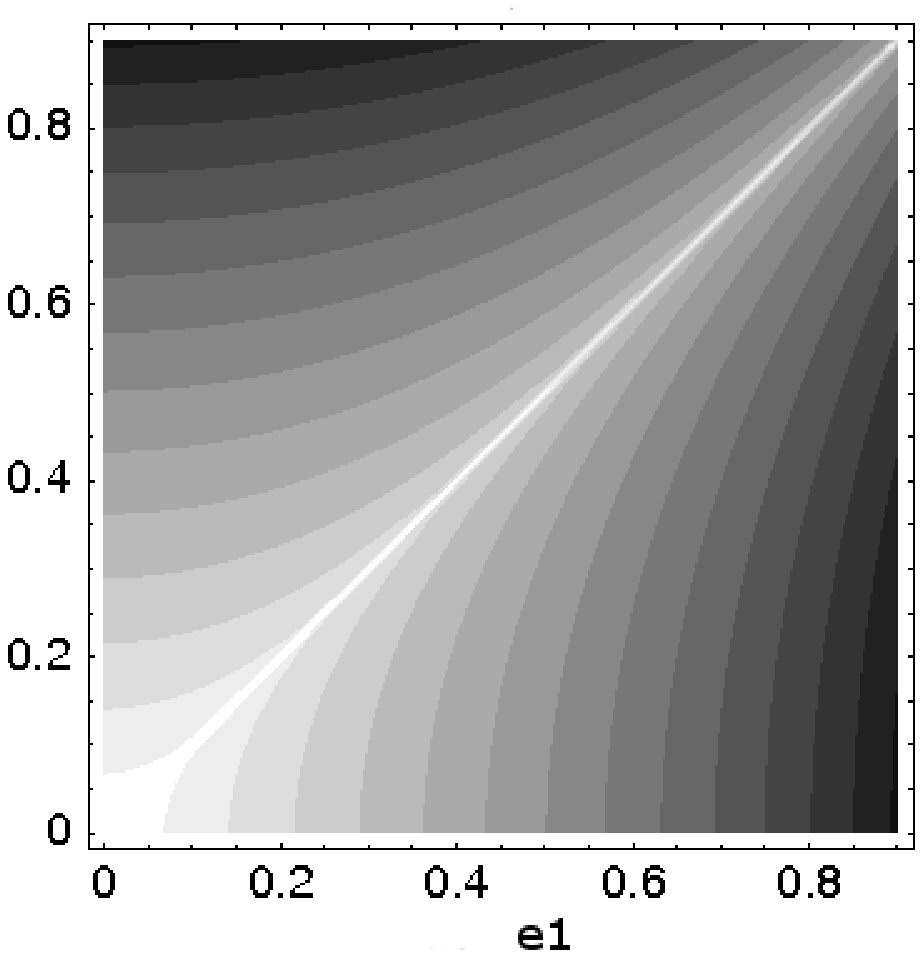}
 \hspace*{-3mm}
 \includegraphics[width=4.5cm,totalheight=3.66cm]{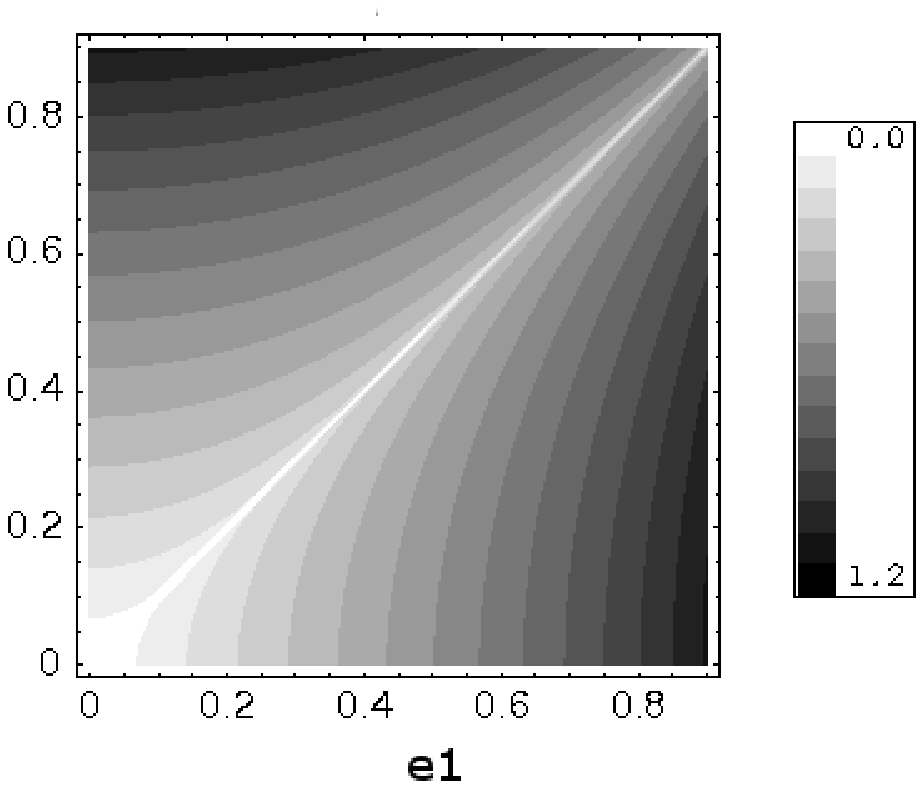}
\end{center}
\caption{ 
Contour plot of collisional velocity $V_{imp}(e_1,e_2)$, in units of $V_{kepler}$,
for the 2:1 resonance (left), 3:2 (middle), 4:3 (right)
with libration width $A=0.1\pi$ (top), $A=0.3\pi$ (middle), and $A=\pi$
(bottom, non-resonance case).
Darker regions correspond to higher velocities.
 \label{contourfig_vel}
 }  
\end{figure*}

In this section, we show numerical results for the collisional velocity
$V_{imp}(e_1,e_2)$, calculated with the aid of \eqref{vimp_e1e2} and  \eqref{deltak_final}.
Recall that this velocity
is the average collisional velocity between two subsets of particles in the resonant family:
one with eccentricity $e_1$ and another one with eccentricity $e_2$.

Figure \ref{fig_vele1} (left column) shows this velocity as a function of one
of its two arguments, with the second argument fixed to 0.1.
The velocity is measured in units of the circular Keplerian velocity,
$V_{kepler} \equiv \sqrt{GM/a}$.
The major effect seen in these plots is that $V_{imp}(e_1,e_2)$ decreases from $e_2$ at $e_1=0$
to zero at $e_1=e_2$ and then increases again. The same $V$-shape pattern is seen in
right panel, resulting from our test numerical integrations, in which we trapped planetesimals
into a 2:1-resonance with a slowly migrating, Neptune-like planet.

The results were obtained using the algorithm developed by \cite{thebault-brahic-1998}
and \cite{thebault-doressoundiram-2003}.

The two left panels show the resonances of different strength, a shallow one at the top and
a strong one at the bottom. Different curves correspond to different resonance numbers $p$ and $q$.
Although the resonant lock does affect the velocities, is it somewhat surprising that
the effect is rather subtle.

Fig. \ref{contourfig_vel} depicts the velocity in the $e_1$-$e_2$-plane.
The upper panels for a strong resonance shows, as expected, a sharp minimum along the line of equal 
eccentricities and an extended area of low collisional velocity where both
eccentricities are low, $e_1,\; e_2 \sim 0 \dots 0.2$.
As far as other regions of the plane are concerned, the effects are highly
non-linear. Particularly interesting is a sharp increase of the collisional velocity
from moderate to large eccentricities that occurs for 3:2 and 4:3 resonances.
It is due to the  fact that, starting from a certain $e$, the resonant ``clumps''
start to overlap, see two rightmost panels in the middle row of Fig.~\ref{garden}.
The larger $p$, the smaller the eccentricity at which the effect shows up.
This is because a $(p+q):p$ resonance produces $p$ clumps;
the larger the number of clumps, the easier they start to overlap.
The same effects, in a weaker form, are seen in the middle panels that are drawn for a
shallower resonance.
The velocity is highest in collisions between particles in highly-eccentric orbits
with those in moderately-eccentric ones.
In the non-resonant case depicted in the lowest panels,
the highest velocity is attained in collisions between particles in highly-eccentric orbits
with those in nearly-circular ones.
The maximum possible value of $V_{imp}(e_1,e_2)$ is $\sqrt{2}$.
It is achieved asymptotically when $e_1=0$ and $e_2 \rightarrow 1$.

\subsection{Average collisional velocity in the disk}

We now calculate the average collisional velocity in the whole disk of particles.
This is accomplished by integrating $V_{imp}(e_1, e_2)$ over both $e_1$ and $e_2$
from $0$ to $e_{max}$, Eq.~\eqref{phie}.
In such a way, we
get the average collisional velocity in the disk as a function of the maximum
possible eccentricity, $V_{imp}(e_{max})$:
\begin{eqnarray}
\begin{split}
V_{imp}&(e_{max}) = \\
                    & \int_{e_1}\int_{e_2} 
                     \frac{\Delta^{(1)}(e_1,e_2)}{\Delta^{(0)}(e_1,e_2)}
		     \phi_e(e_1 ; e_{max}) \phi_e(e_2 ; e_{max})
		     de_1 de_2
\label{vimp_emax}
\mbox{ .}
\end{split}
\end{eqnarray}
The additional double integration is done numerically with a Monte-Carlo method.

\begin{figure}
\begin{center}
\psfrag{emax}{$e_{max}$}
\psfrag{Vemax}{\vspace*{2ex} $V(e_{max})$}
 \includegraphics[totalheight=0.3\textheight]{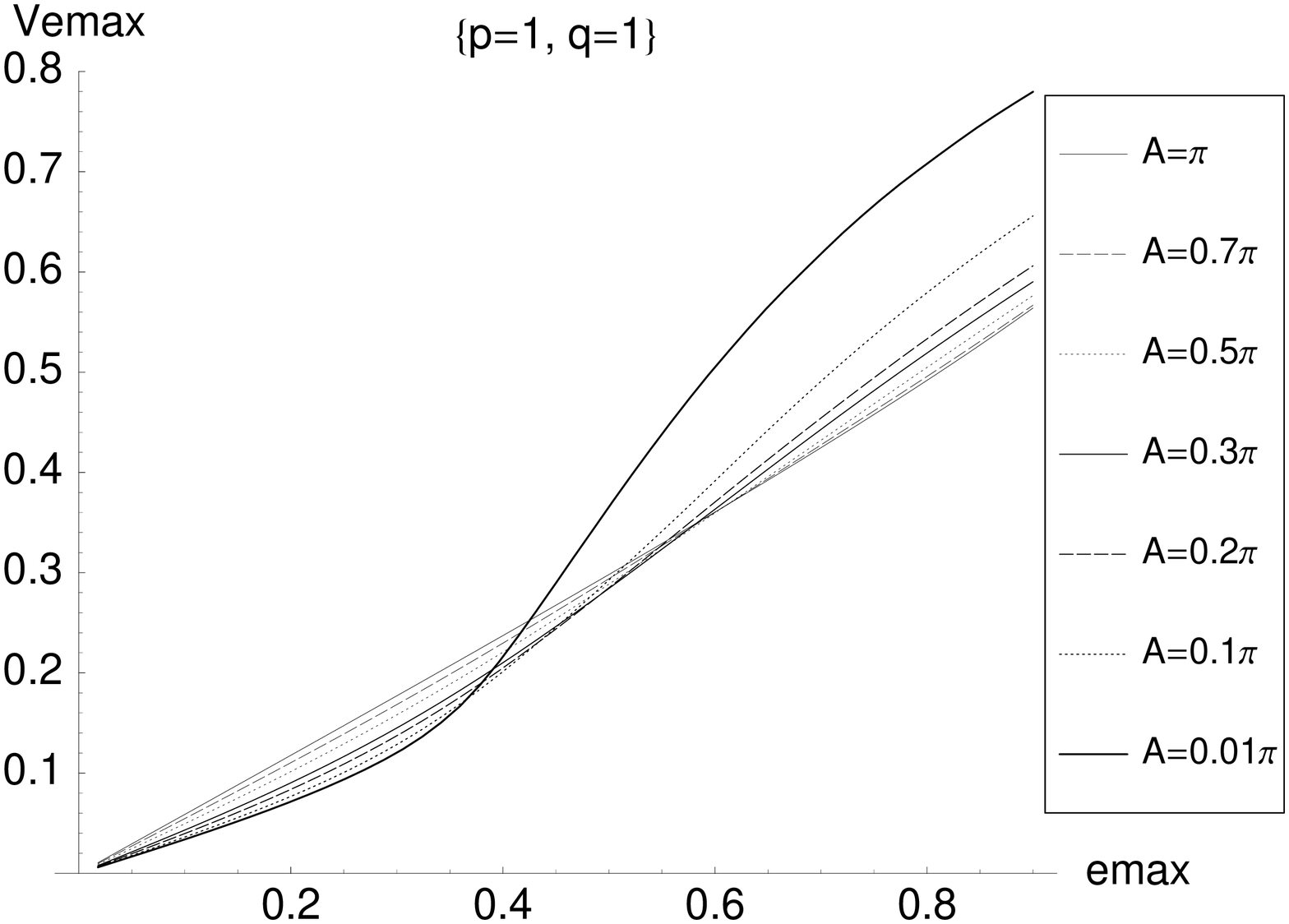} 
 \includegraphics[totalheight=0.3\textheight]{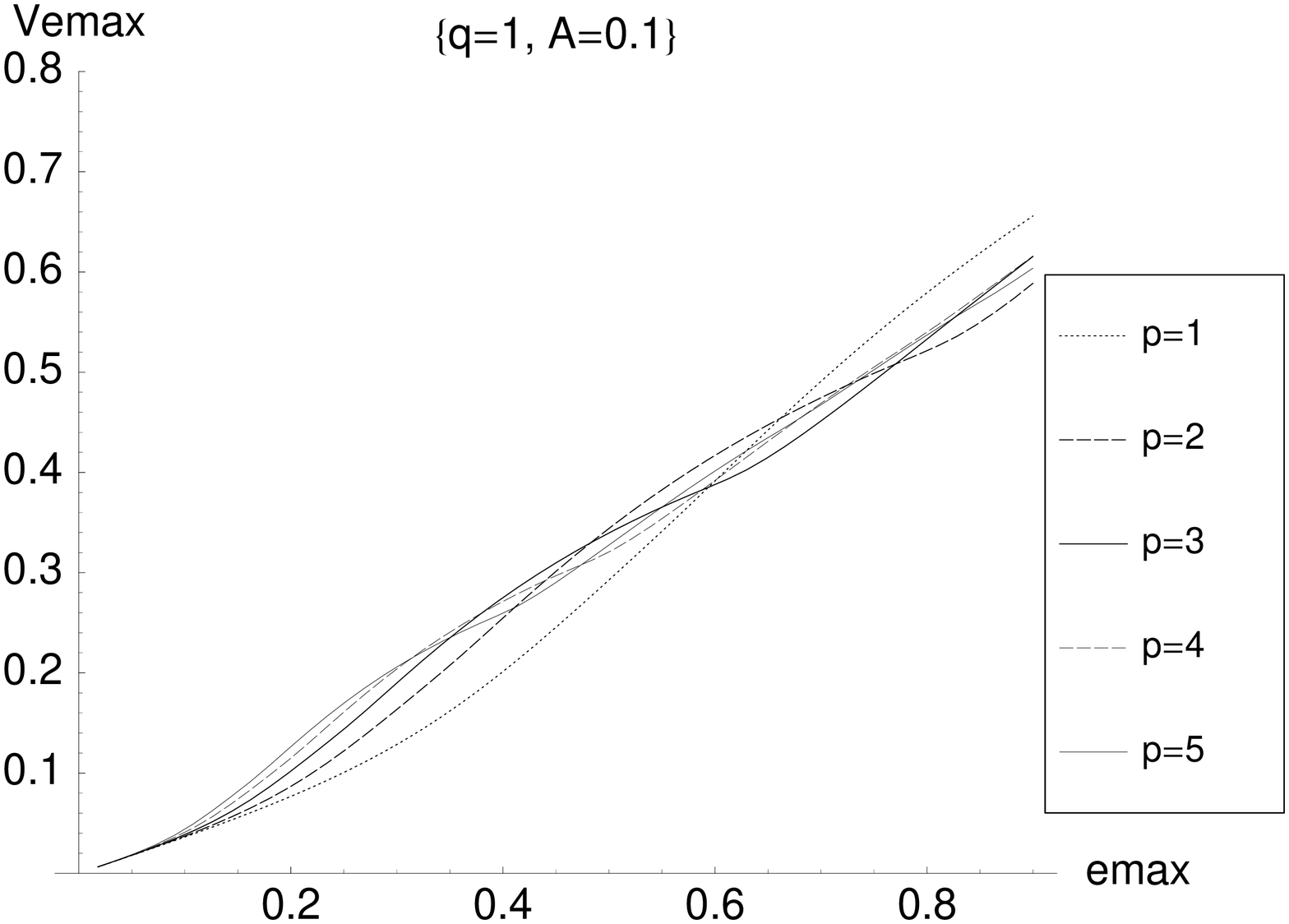} 
 \includegraphics[totalheight=0.3\textheight]{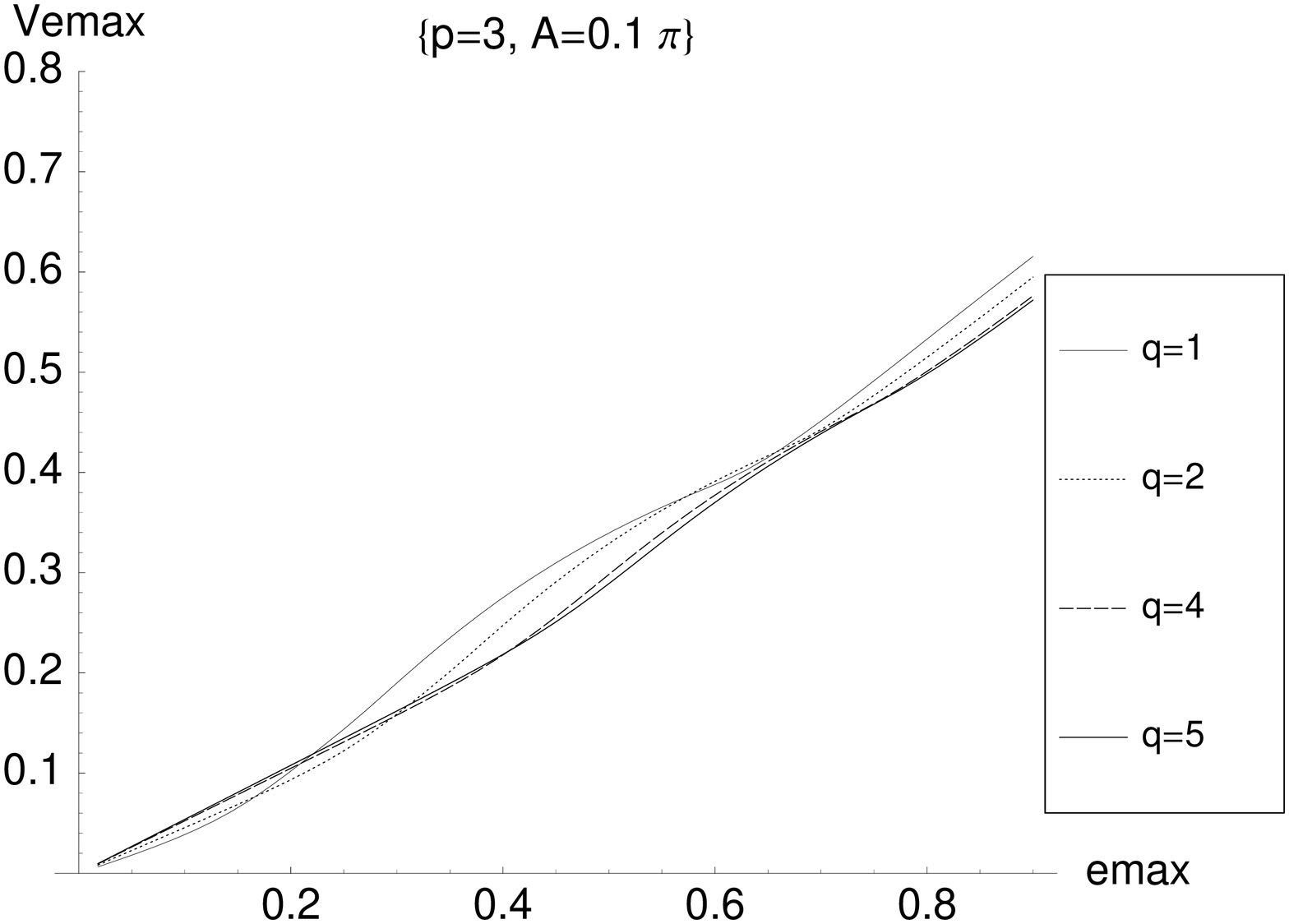} 
\end{center}
\caption{ 
 Collisional velocity $V_{imp}(e_{max})$ in units of $V_{kepler}$.
 Top: dependence on libration width $A$ width for the 2:1 resonance.
 Middle: dependence on resonance parameter $p$ for first-order resonances $q=1$, $A=0.1\pi$.
 Bottom: dependence on the order of resonance $q$ for $p=3$, $A=0.1\pi$.
 \label{fig_velemax}
 }  
\end{figure}

Fig. \ref{fig_velemax} (top) shows the dependence on the libration width of the resonant
argument. The smaller $A$, the stronger the resonance -- and its influence -- gets.
$A=\pi$ describes a non-resonant case, whereas $A=0$ corresponds to a perfect resonant lock.
Again, the collisional velocity is affected by the resonance only weakly, but the
effect is present.
It is interesting that for moderate eccentricities, the average collisional velocity in a resonant belt 
is lower than in a similar non-resonant one, while for large eccentricities, the opposite is true.
Similarly,  middle and bottom panels in Fig.~\ref{fig_velemax} demonstrate the dependence
of $V_{imp}$ on $p$ and $q$. They show that collisional velocity does vary with the resonant integers,
and may be both higher or lower that the non-resonant one, but the effect is usually
of the order of several tens of percent.

Remember that the whole treatment of the collisional velocities presented above
has been done purely in 2D. Therefore, in real applications,
the results must be carefully corrected for inclination terms 
($i \, V_{kepler}$) in velocity.

%-----------------------------------------------------------------------
\section{Collisional rates and collisional lifetimes}
%-----------------------------------------------------------------------
  
Using the same methods as in the previous section,
we will now investigate the influence of a resonance on the frequency
of collisions.
One might expect
the number density of the particles in the resonant ``clumps'', and therefore
the collisional rate, to be higher than
in a similar non-resonant belt.
We check whether, and to what extent, this expectation is true.

\subsection{Collisional rate for the subsets of particles with $e=e_1$ and $e=e_2$}

Like in the case of collisional velocity, we start by considering
two subsets of particles in the resonant family:
one with eccentricity $e_1$ ($n_1$ particles)
and another one with eccentricity $e_2$ ($n_2$ particles)
and ``count'' collisions between particles of population~1 with those of population~2.
The rate of collisions is given by
\citep{krivov-et-al-2005,krivov-et-al-2006},
\begin{equation}
  R(e_1,e_2)
  =
  N \sigma
  V_{imp}(e_1, e_2) \Delta^{(0)}(e_1, e_2)
\label{R2Dinitial}
\end{equation}
which simplifies to
\begin{equation}
  R(e_1,e_2)
  =
  N \sigma
  \Delta^{(1)}(e_1, e_2) .
\label{R2D}
\end{equation}
Here, 
$\sigma$ is the collisional cross-section for equal-sized particles,
and the front factor $N$ depends on what we mean by ``collisional rate''.
If $R(e_1, e_2)$ is the number of collisions per unit time that a particle
of population~1 has with any particle of population~2, then $N=n_2$.
Conversely, if we consider a particle of population~2 colliding with
population 1 particles, then $N=n_1$.
Finally, to get the total number of collisions between particles of both
families occurring per unit time, we should set $N=n_1 n_2$.

However, the above formulas are two-dimensional. For instance,
$\sigma$ should be understood as $2s$, where $s$ is the radius of
equal-sized particles.
Any physically meaningful calculation of collisional frequencies
requires a 3D treatment.
We thus introduce an approximation for a 3D $\Delta$-integral,
\begin{equation}
\Delta^{(k)}_{3D}(.)
\equiv
{ \Delta^{(k)}(.) \over h},
\label{3D}
\end{equation}
where
$h = 2 r \sin\epsilon \approx 2 a_{res} \sin\epsilon$
is the disk thickness at the annulus location,
$\epsilon$ being the half-opening angle of the disk.
Thus the 3D collision rate is
\begin{equation}
  R(e_1,e_2)
  =
  N \sigma
  \Delta^{(1)}_{3D}(e, e_2),
\label{R_def}
\end{equation}
now with the ``usual'' collisional cross section
$\sigma = 2 \pi s^2$.
We note that 
 $\Delta^{(1)}_{3D}$, i.e.
$R$ without the factor $N \sigma$, is exactly what is usually called
``intrinsic collisional probability'' \citep[e.g.][]{greenberg-1982,davis-farinella-1997}.

\begin{figure*}
\psfrag{Ree}{\hspace*{-3ex} $\tilde{R}(e_1,e_2=0.1)$}
\psfrag{e1}{$e_1$}
\begin{tabular}{cc}
 \includegraphics[totalheight=0.25\textheight]{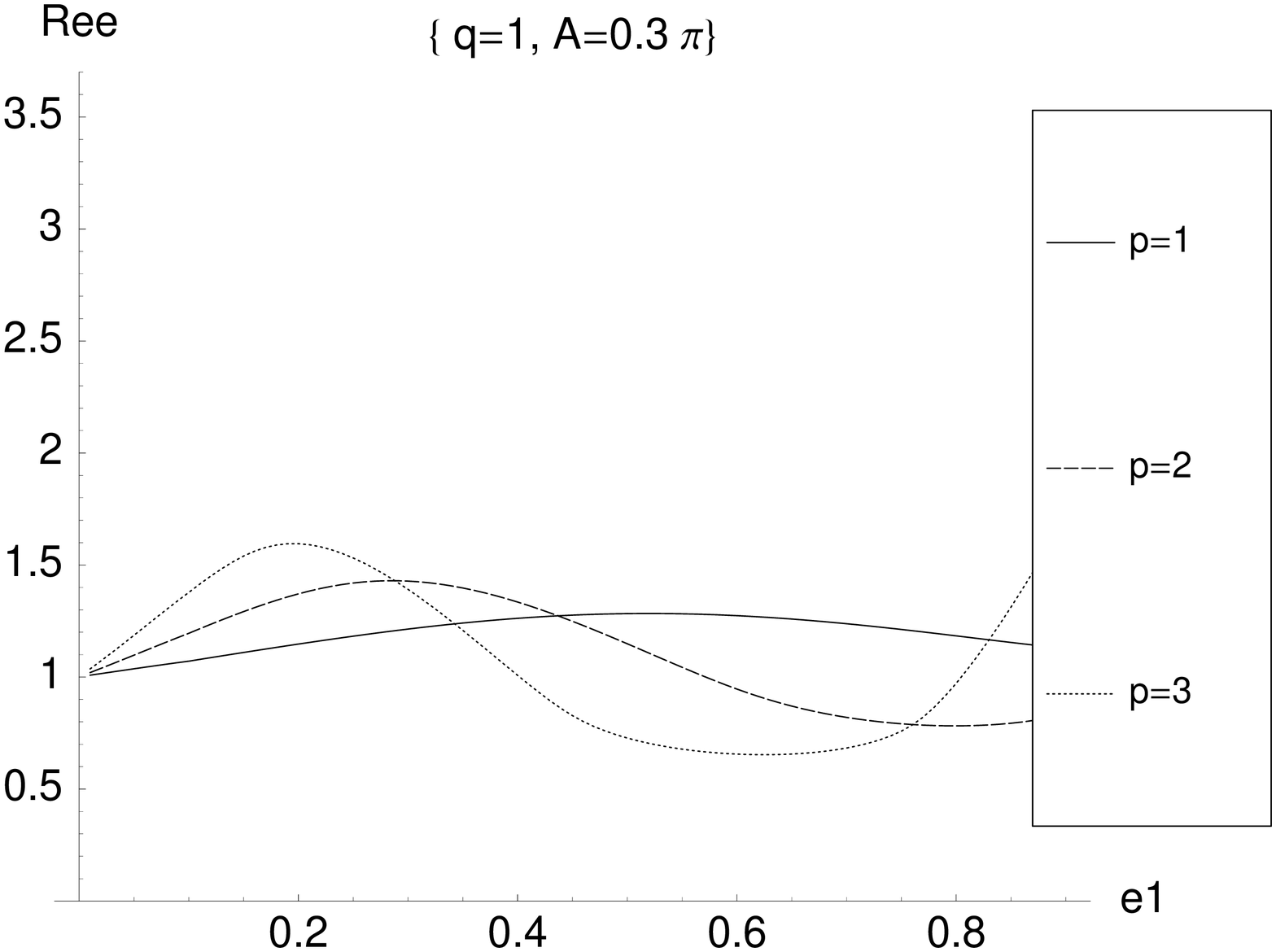}
 & \\
 \includegraphics[totalheight=0.25\textheight]{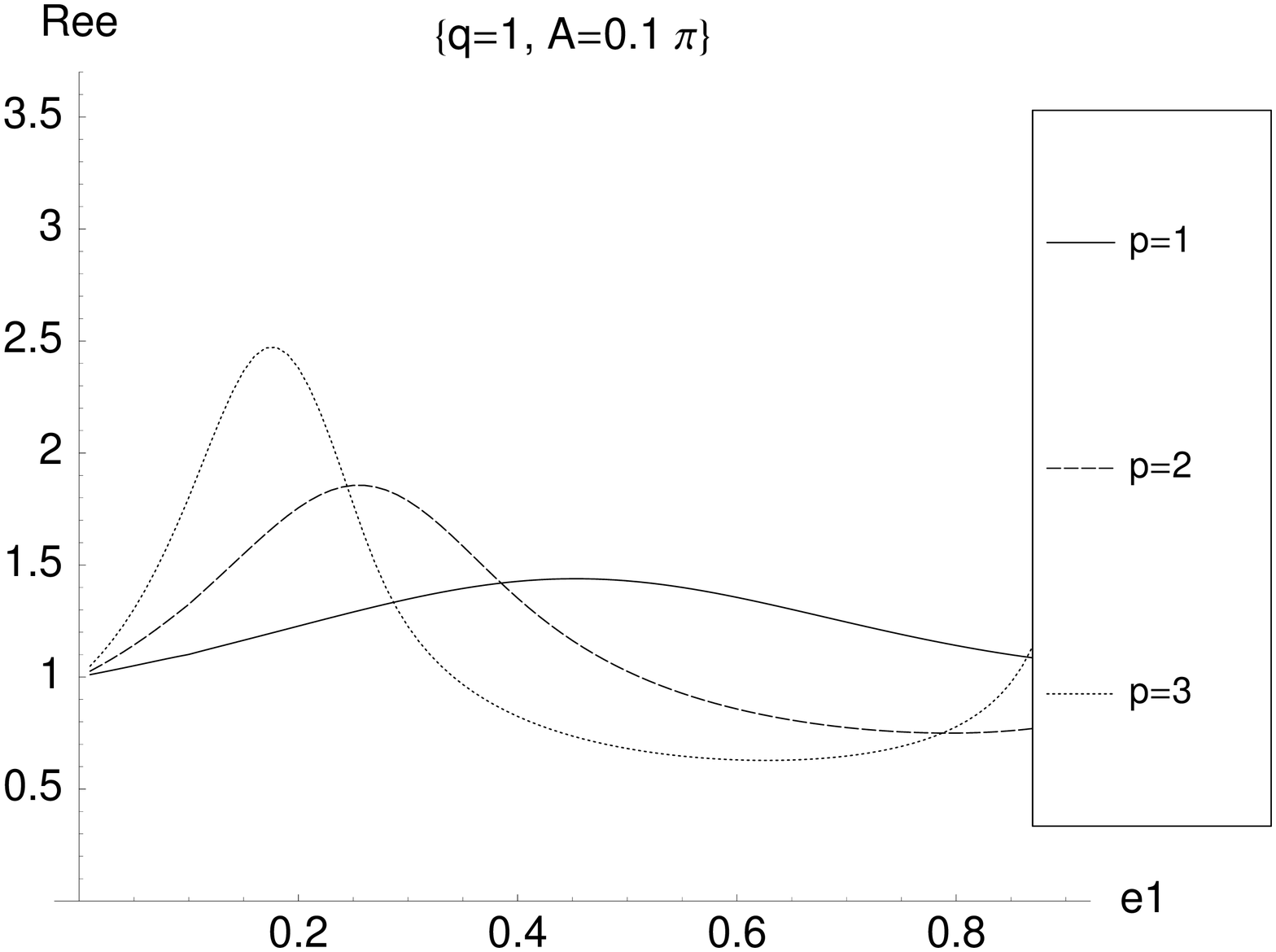}
 &
 \includegraphics[totalheight=0.17\textheight]{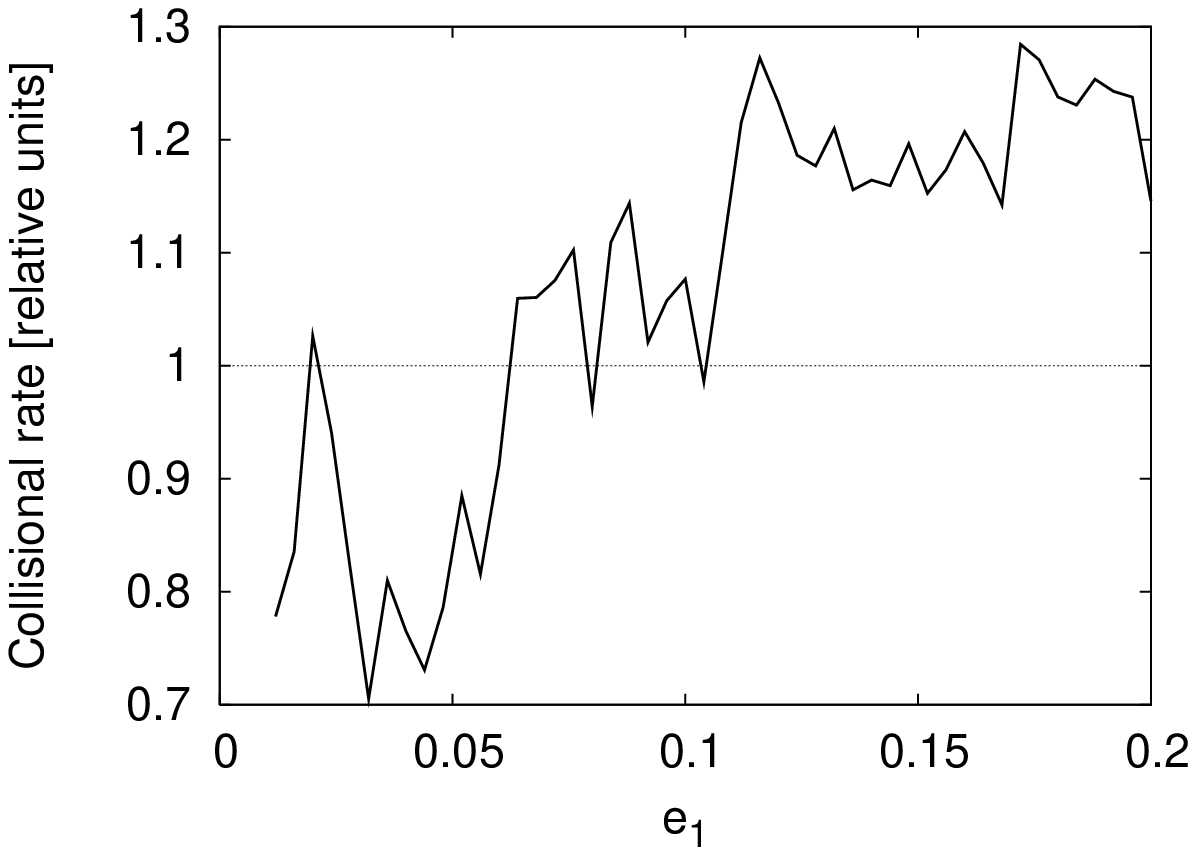}
\end{tabular}
 \caption{
 Cuts through the contour plots of collisional rate $\tilde{R}(e_1,e_2=0.1)$,
 for the 2:1 resonance (solid), 3:2 (dashed), 4:3 (dotted).
 Top left: a shallower resonance with $A=0.3\pi$, 
 bottom left: a strong one with $A=0.1\pi$.
 Right panel:
 typical numerical results
 for $R(e_1,e_2=0.1)$ in a 2:1 resonance for comparison.
 \label{fig_ratee1}
 }
\end{figure*}

\begin{figure*}[htb!]
\begin{center}
\psfrag{e2}{$e_2$}
\psfrag{top}{}
\psfrag{bottom}{}
 \includegraphics[totalheight=3.5cm]{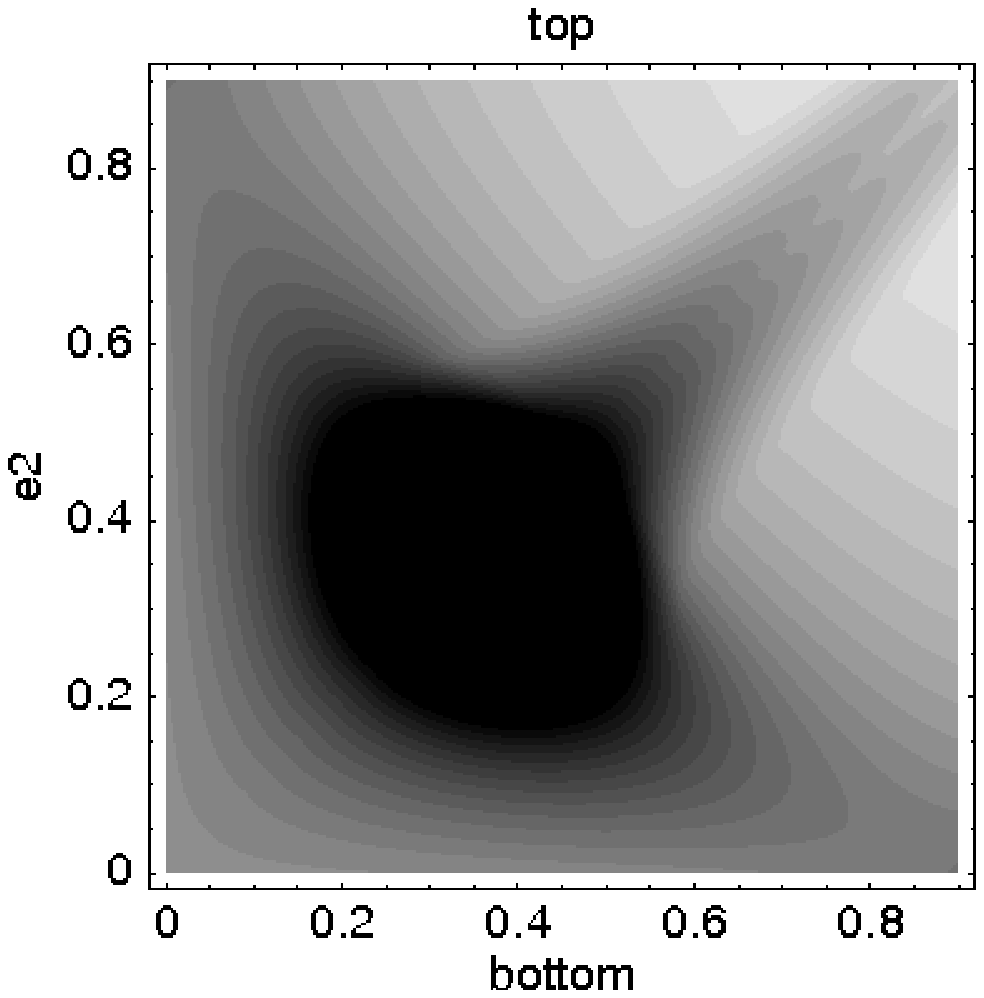}
 \includegraphics[totalheight=3.5cm]{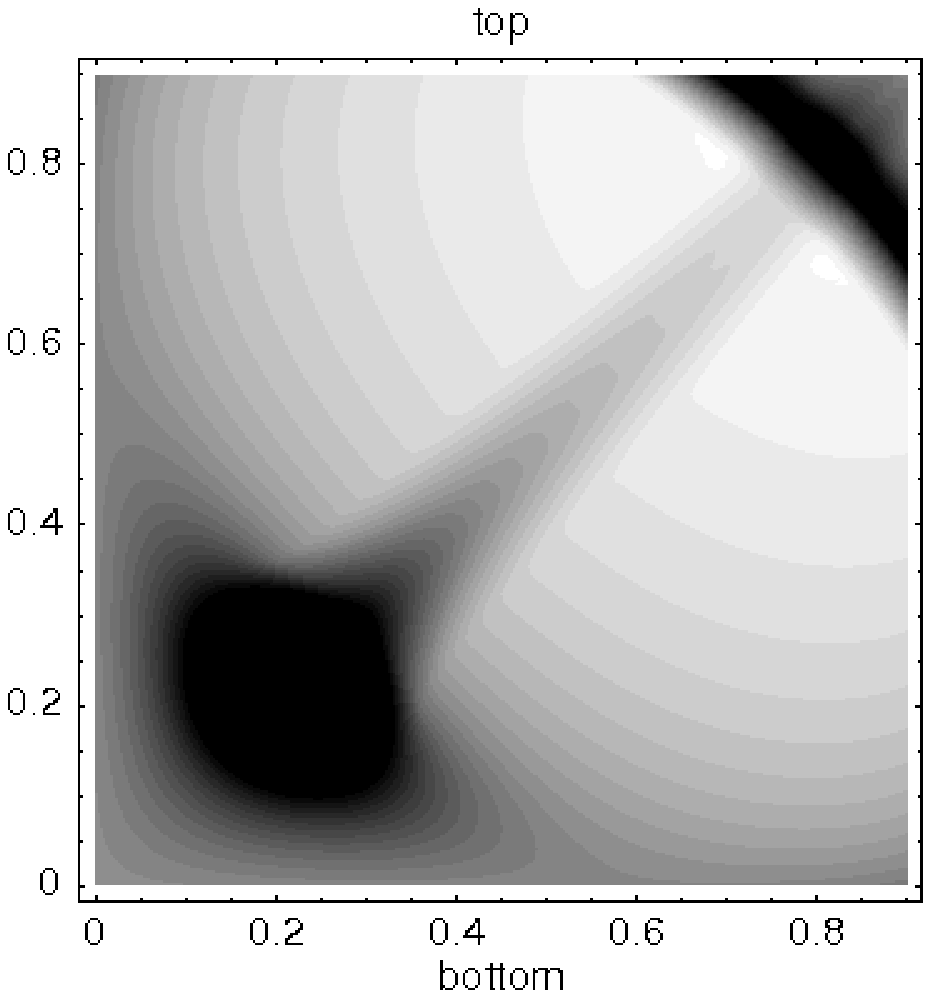}
 \hspace*{-3mm}
 \includegraphics[totalheight=3.66cm]{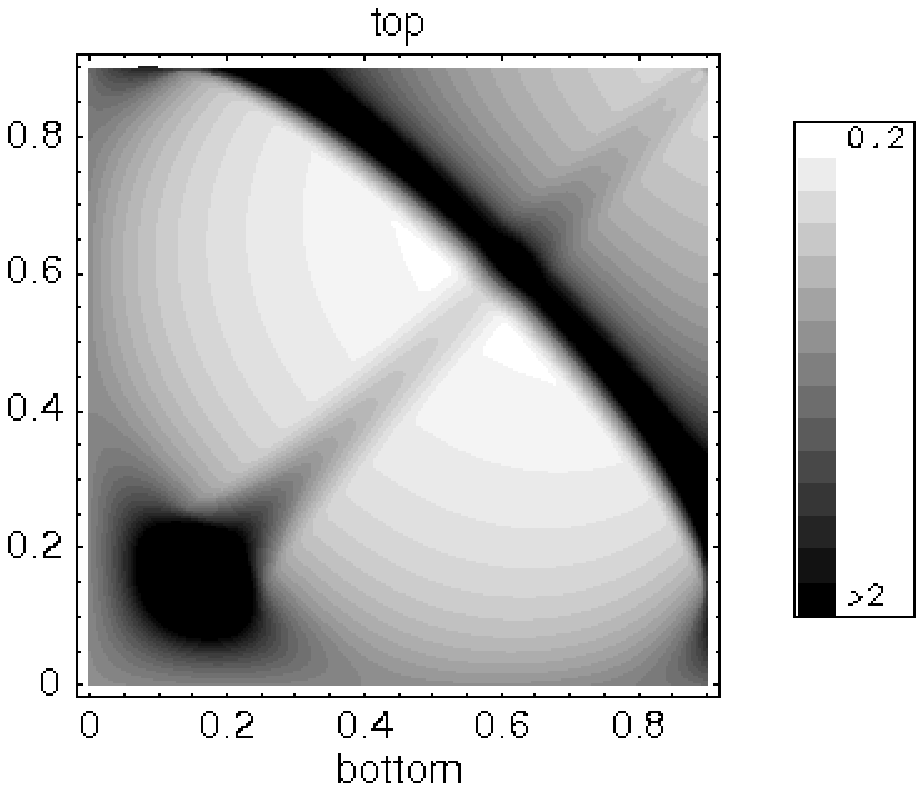}\\[-3mm]
\psfrag{top}{}
\psfrag{bottom}{}
 \includegraphics[totalheight=3.5cm]{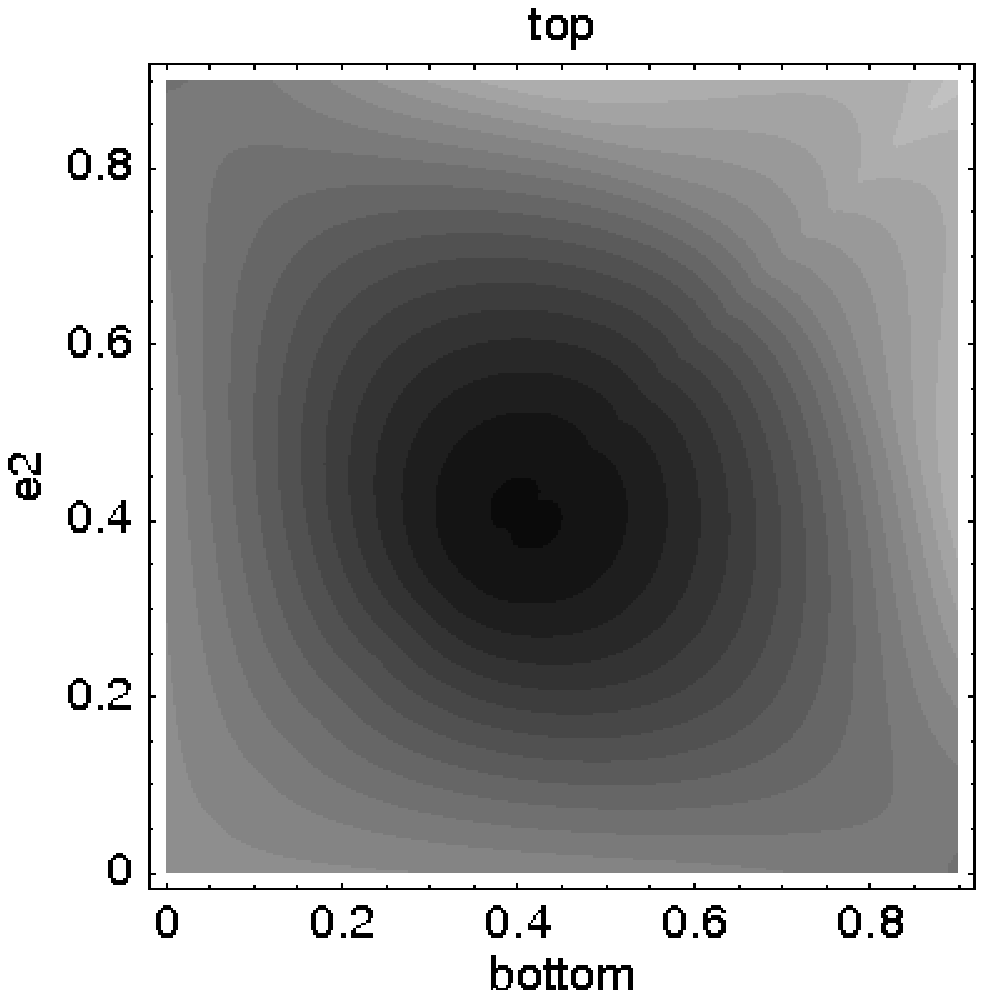}
 \includegraphics[totalheight=3.5cm]{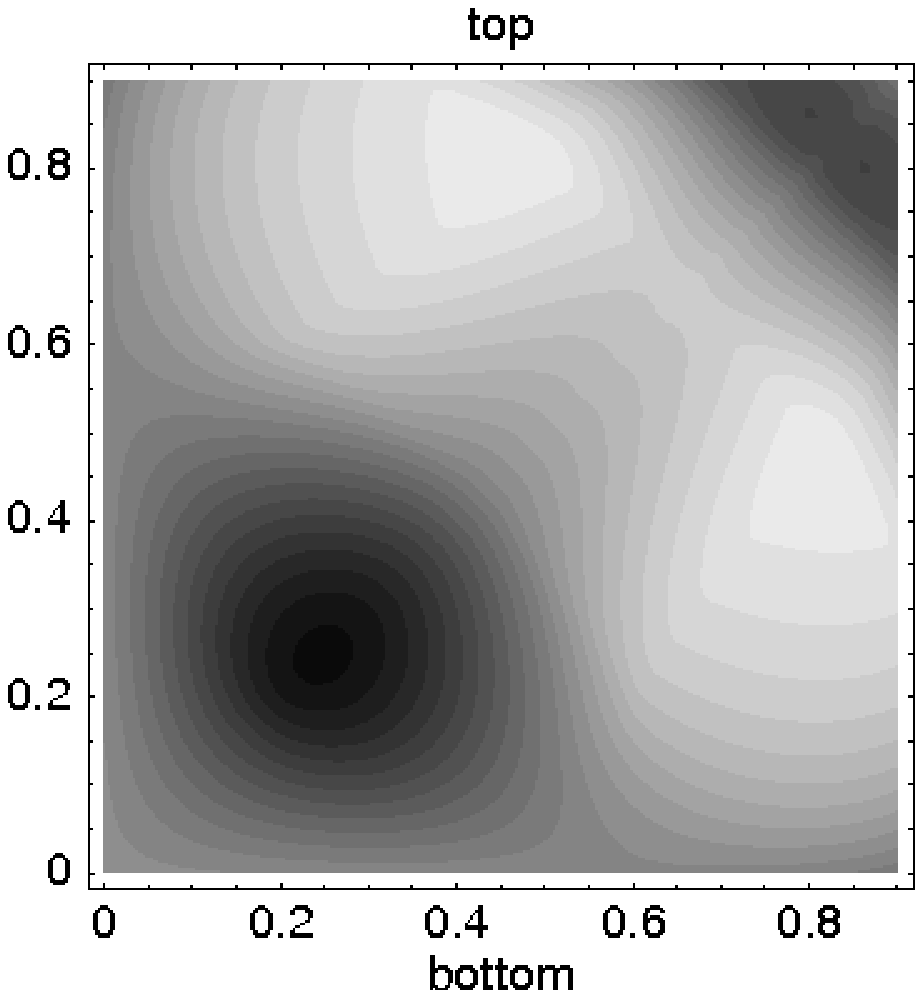}
 \hspace*{-3mm}
 \includegraphics[totalheight=3.66cm]{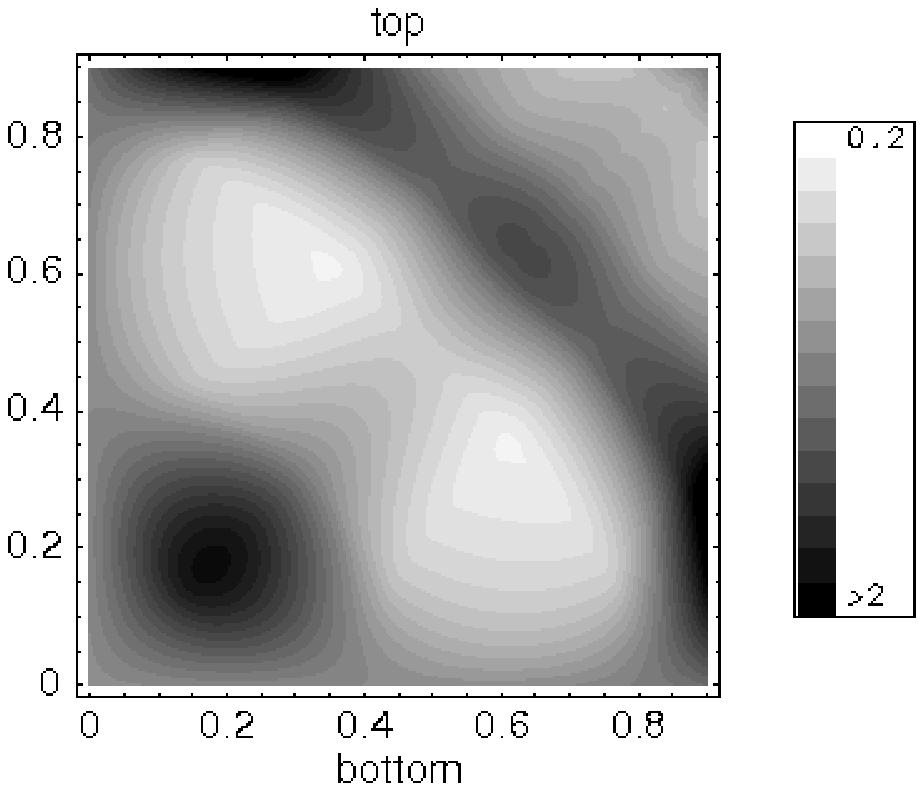}\\[-3mm]
\psfrag{top}{}
\psfrag{bottom}{\hspace*{1.3ex} $e_1$}
 \includegraphics[totalheight=3.5cm]{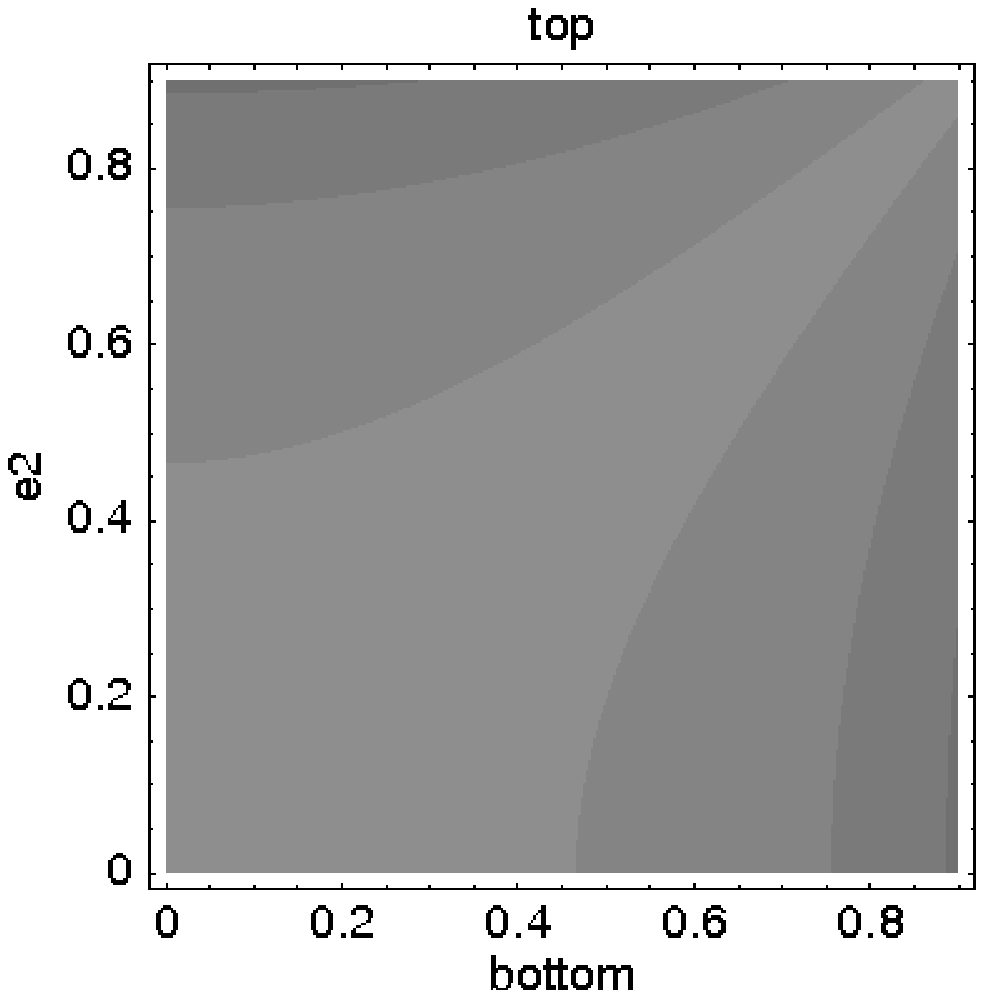}
 \includegraphics[totalheight=3.5cm]{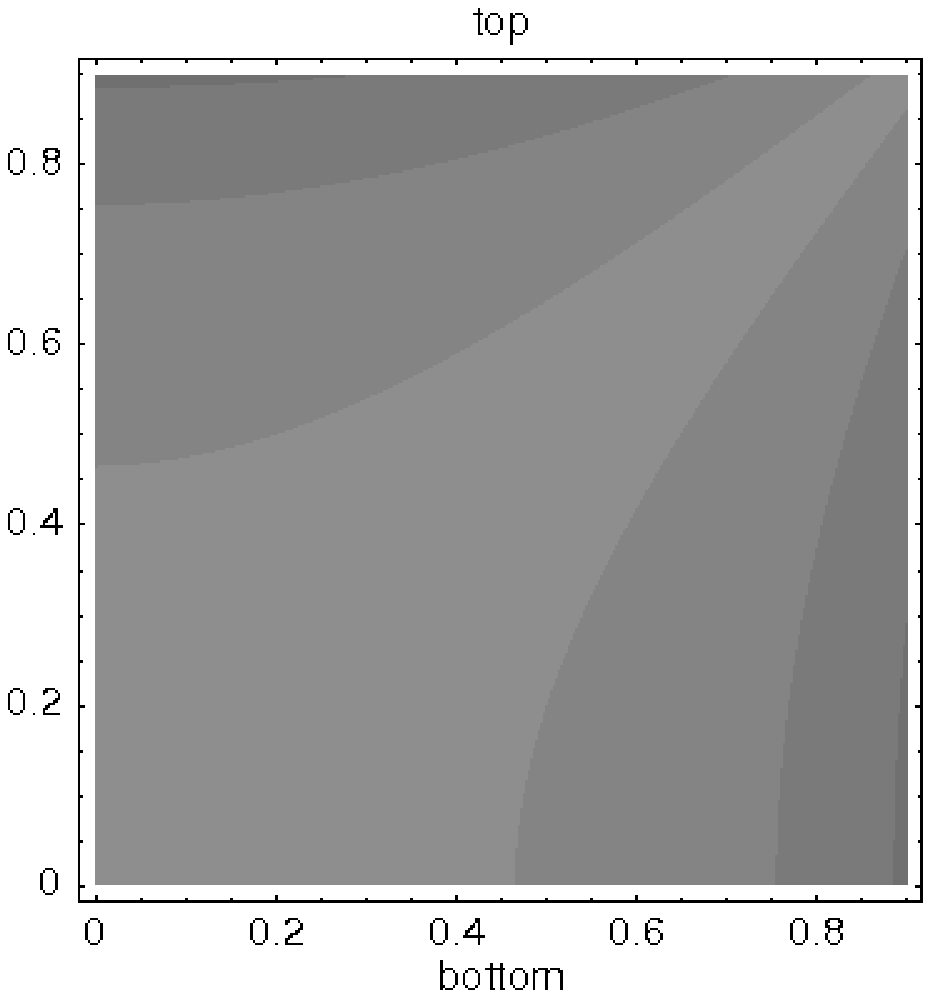}
 \hspace*{-3mm}
 \includegraphics[totalheight=3.66cm]{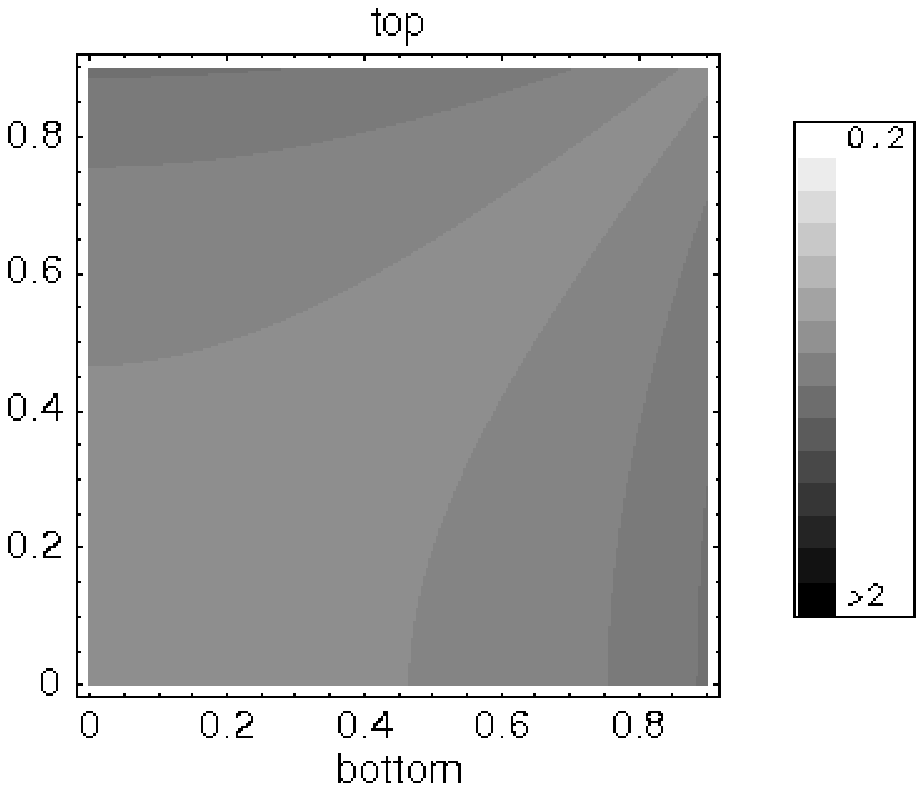}
\end{center}
\caption{ 
Contour plot of the dimensionless collisional rate $\tilde{R}(e_1,e_2)$,
for the 2:1 resonance (left), 3:2 (middle), 4:3 (right)
with libration width $A=0.1\pi$ (top), $A=0.3\pi$ (middle), and $A=\pi$
(bottom, non-resonance case).
The darker the contours, the higher the rate.
 \label{contourfig_rate}
 }  
\end{figure*}

Instead of dealing with the full collisional rate \eqref{R_def}, 
it is convenient to introduce a dimensionless factor $\tilde{R}(e_1,e_2)$:
%which changes with the resonance parameters $p,\;q,\;A$:
\begin{equation}
  \tilde{R}(e_1,e_2) 
  \equiv
  R(e_1,e_2) / R_0  
,
\label{Rtilde}
\end{equation}
where
\begin{equation}
  R_0 \equiv
  {1 \over 2\pi^2}
  {N \sigma V_{kepler}  \over h \; a_{res}^2} 
\label{R_0}
\end{equation}
is the approximate ``particle-in-a-box'' value of the collisional 
rate\footnote{See, e.g.,
Eq. (18) in \citet{krivov-et-al-2006b}.
Note that in Eqs. (15), (17), and (18) of that paper, an additional factor 1/2
is erroneously missing.
}. %end footnote 
In the first order, $R_0$ does not depend on eccentricities
of the colliding particles and depends on their
inclinations through the divisor $h$ only.
It describes the rate of collisions in a non-resonant,
and therefore rotationally-symmetric,
ring of objects.
Again, $R_0/(N \sigma)$ is the intrinsic collisional probability
in such a ring.
Therefore the deviations of the dimensionless collisional rate $\tilde{R}$ 
from unity would describe effects of the resonance, as well as some corrections due
to non-zero eccentricities.

Figure \ref{fig_ratee1} (left column) shows the
dimensionless collisional rate $\tilde{R}(e_1,e_2)$
as a function of one of its two arguments,
with the second argument fixed to 0.1.
For large $A$, i.e. for a weak resonance, shown in the top left panel,
the collisional rate is almost independent of $e_1$, being close to the ``particle-in-a-box''
value.
For stronger resonances (bottom left panel) $\tilde{R}(e_1,e_2)$ peaks at intermediate values of 
eccentricity $e_1$.
The stronger the resonance, the more pronounced the maximum.
For the 4:3 resonance and $A=0.1\pi$, the maximum collisional rate 
at $e_1\sim 0.2$ is about 2.5 times larger than the non-resonant rate.
Interestingly, the larger the resonant integer $p$, the smaller the value of 
eccentricity, at which the collisional rate is the highest.
Another finding is that, for sufficiently large eccentricities, resonances may decrease
the frequency of collisions.
For comparison,
in the right panel we show a typical result of our numerical integrations, in which we trapped 
planetesimals into a 2:1-resonance with a slowly migrating planet.
Notwithstanding large oscillations, caused by a limited number of particles integrated,
it shows the same tendency of the collisional rate to slightly increase with $e_1$.

Fig. \ref{contourfig_rate} depicts the collisional rate in the $e_1$-$e_2$-plane.
The upper and middle panels show that collisions are most frequent between particles,
whose orbital eccentricities are moderate and not very different from each other.
Interestingly, these are exactly the eccentricities for which collisional rates in the non-resonant 
case are the lowest (bottom panels).
For some resonances, another region of higher collisional rates is observed at very high 
eccentricities.
This effect has the same origin as a similar effect in the collisional velocities,
Fig.~\ref{contourfig_vel}.

\subsection{Average collisional rate in the disk}

\begin{figure}
\begin{center}
\psfrag{emax}{$e_{max}$}
\psfrag{RR0}{\hspace*{-2ex} $\tilde{R}(e_{max})$}
 \includegraphics[totalheight=0.3\textheight]{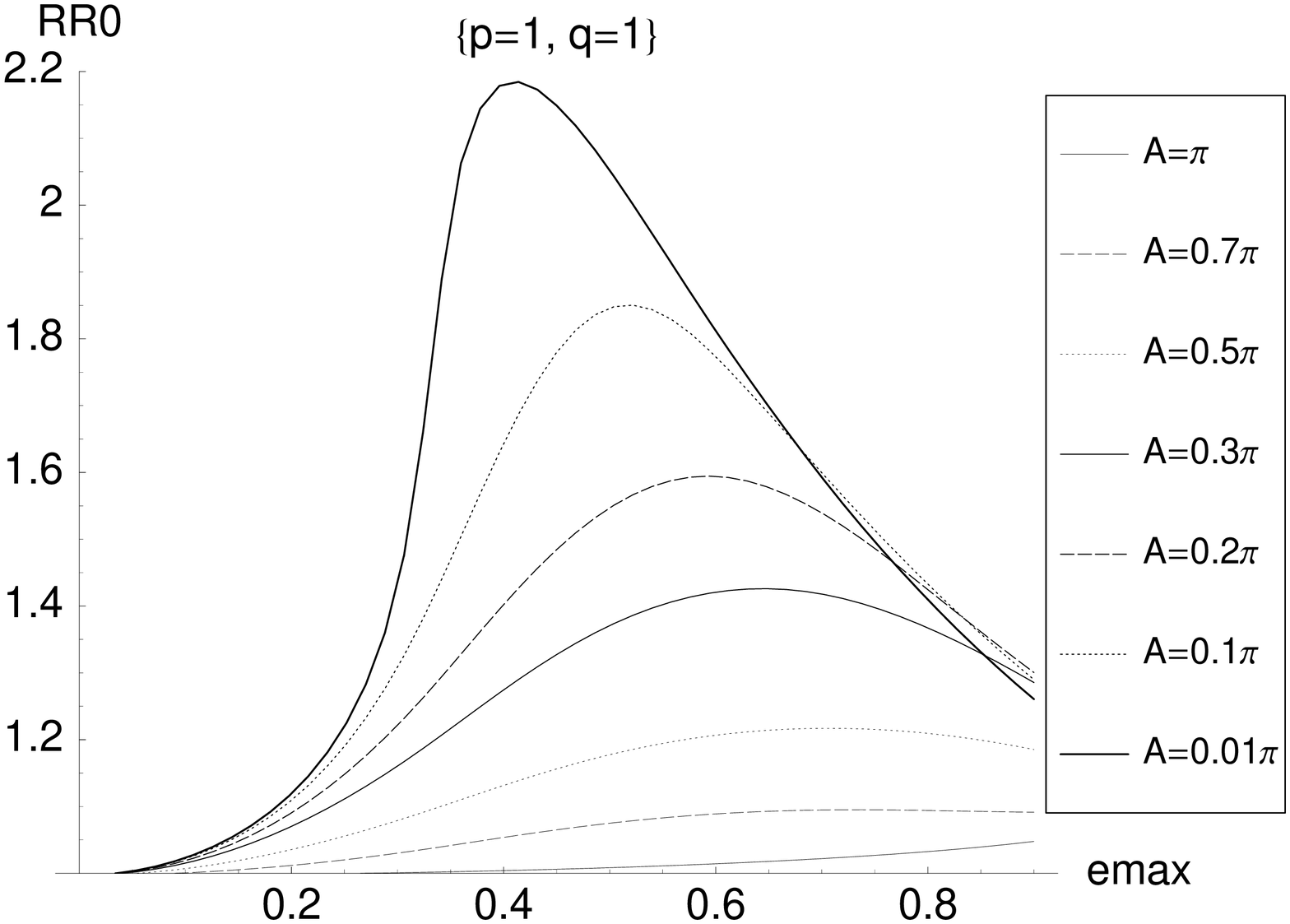} 
 \includegraphics[totalheight=0.3\textheight]{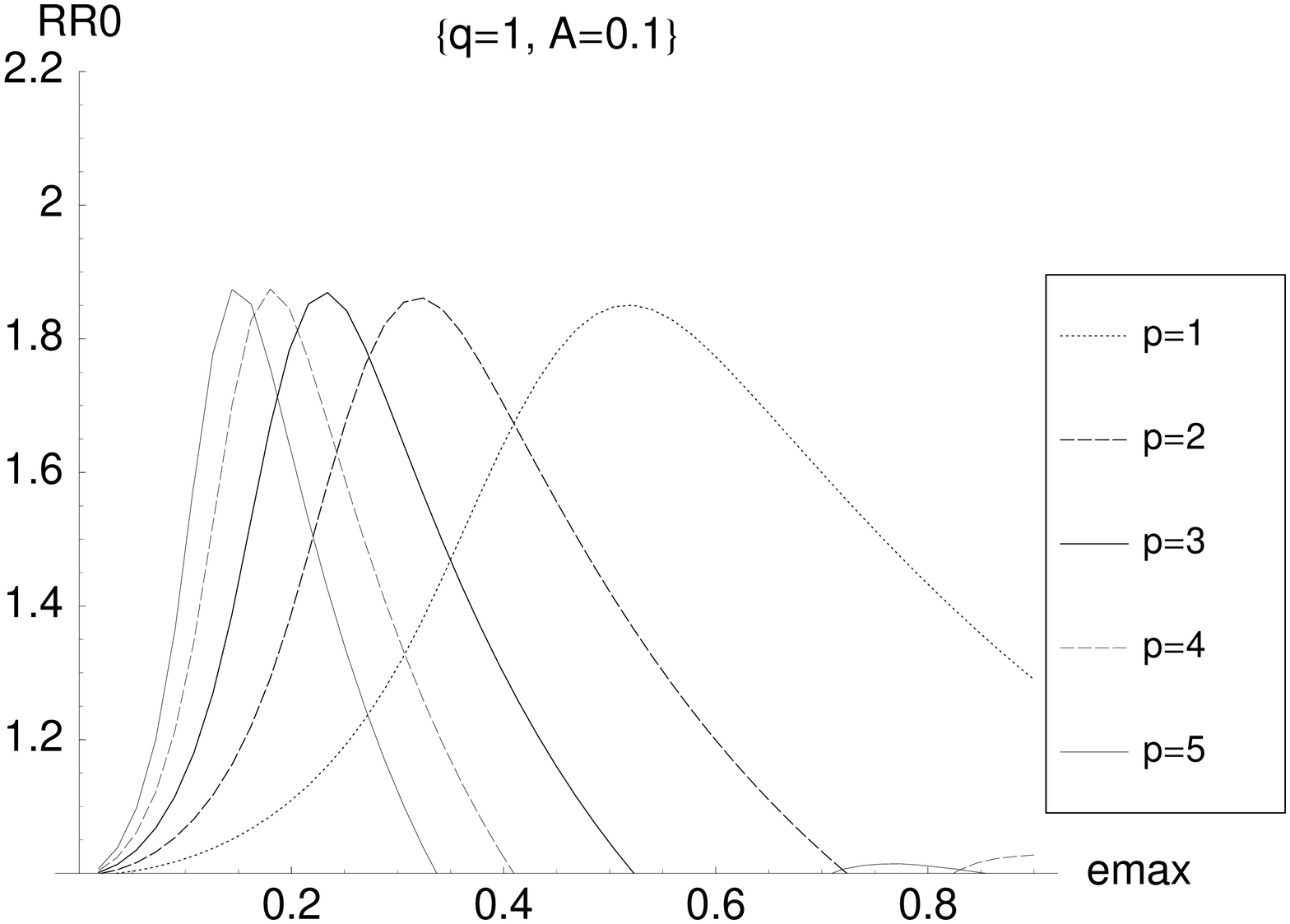} 
 \includegraphics[totalheight=0.3\textheight]{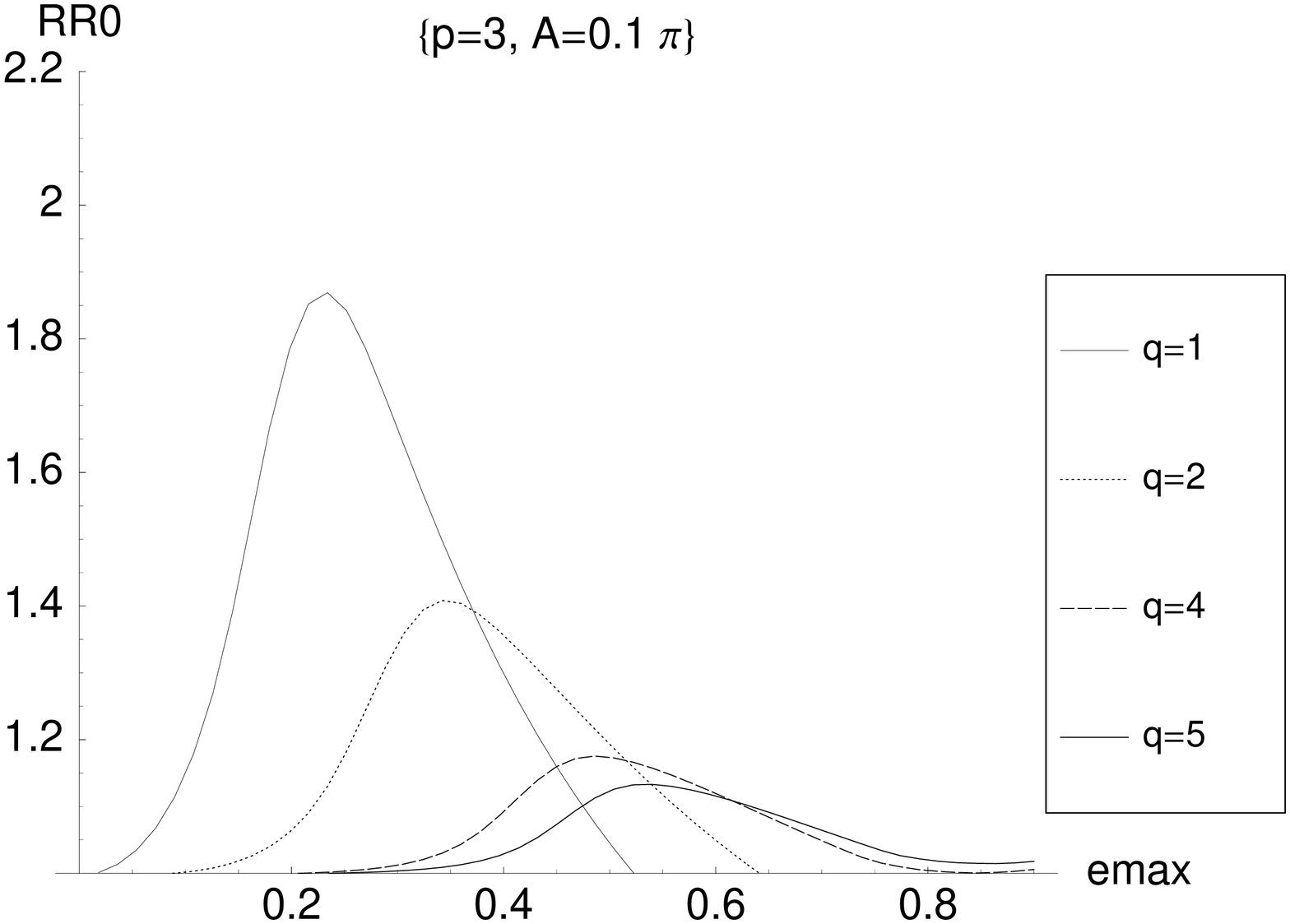} 
\end{center}
 \caption{ 
 Dimensionless collisional rate $\tilde{R}(e_{max})$ in the disk,
 Top: dependence on libration width $A$ width for the 2:1 resonance.
 Middle: dependence on resonance parameter $p$ for first-order resonances $q=1$, $A=0.1\pi$.
 Bottom: dependence on the order of resonance $q$ for $p=3$, $A=0.1\pi$.
 \label{fig_rateemax}
 }  
\end{figure}

Like in the case of collisional velocity, we now average over both eccentricities
in order to obtain the collisional rate in the whole disk containing $n$ objects:
\begin{eqnarray}
 \begin{split} 
 R&(e_{max})
  = 
\\
&  N \sigma
  \int_{e_1} \int_{e_2} 
  \Delta^{(1)}_{3D}(e_1,e_2)
  \phi_e(e_2;e_{max}) \phi_e(e_1;e_{max})
  de_2 de_1
\label{remax}
\mbox{. }
\end{split}
\end{eqnarray}
Again, the front factor $N$ depends on what we intend to describe with the function $R(e_{max})$.
If $R(e_{max})$ is the number of collisions per unit time that a certain particle
may have with all other particles in the belt, then $N=n$.
To get the total number of all collisions occurring in the ring per unit time,
we should set $N=n^2/2$.

Figure \ref{fig_rateemax} shows the numerical results.
The top panel focuses on the influence of the libration width $A$, whereas
the middle and bottom ones illustrate
the dependence on $p$ and $q$.
It is seen that $R(e_{max})$ has qualitatively the same properties as $R(e_1, e_2)$:
independence of $e_{max}$ out of resonance and a maximum at intermediate eccentricities
for deeper resonances.
The maxima shift to slightly lower eccentricities, when either the libration width decreases,
$p$ increases, or the order of resonance $q$ decreases.
All the curves are generally flatter than the non-averaged ones.

The properties described above can also be understood analytically.
Consider, for instance, the eccentricity $e_\star$ which yields the highest collisional rate,
i.e. the position of the maxima in Fig.~\ref{fig_rateemax}.
By calculating the limit of the $\Delta^{(1)}$-integral at $A \rightarrow 0$
and $e_1, e_2 \rightarrow e$ (see Appendix), one gets
an approximate equation
\begin{equation}
{\sqrt{1 + e_\star} \over (1 -e_\star)^{3/2} }  = { p + q \over p}.
\label{e_star}
\end{equation}
For example, $(p,q)=(1,1)$ gives $e_\star = 0.31$;
$(p,q)=(2,1)$ gives $e_\star = 0.19$;
$(p,q)=(3,1)$ gives $e_\star = 0.14$;
$(p,q)=(3,2)$ gives $e_\star = 0.24$.
A comparison with the positions of relevant maxima in Fig.~\ref{fig_rateemax}, calculated
for a finite $A$, shows that the former lie slightly to the left from the former,
as expected.

The collisional rates are intimately connected to collisional (life)times
of the disk particles.
The average collisional lifetime of an object in the disk is
\begin{eqnarray}
%&& 
T_{coll}(e_{max})
%\nonumber\\
&=&
\left[
  \int_{e_1} \int_{e_2} 
  R(e_1, e_2)
  \phi_e(e_2;e_{max}) \phi_e(e_1;e_{max})
  de_2 de_1
    \over
  \int_{e_1} \int_{e_2} 
  \phi_e(e_2;e_{max}) \phi_e(e_1;e_{max})
  de_2 de_1
\right]^{-1}
\end{eqnarray}
or simply
\begin{eqnarray}
T_{coll}(e_{max}) &=& {1 \over R(e_{max})} ,
\label{Tcoll}
\end{eqnarray}
where $R(e_{max})$ is given by Eq.~\eqref{remax} with $N = n$.

Remember that we have calculated $R$ in quasi-3D. More exactly:
in the collisional rates, which are 
$\propto$ velocity $/$ volume,
we do take into account the inclinations when calculating the volume, but still ignore
those in velocity.

Non-inclusion of the term $\sim i \, V_{kepler}$ in velocity introduces
an error of the same order of magnitude as
in the case of collisional velocities considered in Section~5
\citep[see discussion around Eq. (A14) and (A15) in][]{krivov-et-al-2006}.

%-----------------------------------------------------------------------
\section{The 1:1 resonance: Trojans}
%-----------------------------------------------------------------------

\begin{figure}
\begin{center}
\psfrag{emax}{$e_{max}$}
\psfrag{Vemax}{$V(e_{max})$}
\psfrag{RR0}{\hspace*{-2ex} $\tilde{R}(e_{max})$}
 \includegraphics[totalheight=0.35\textheight]{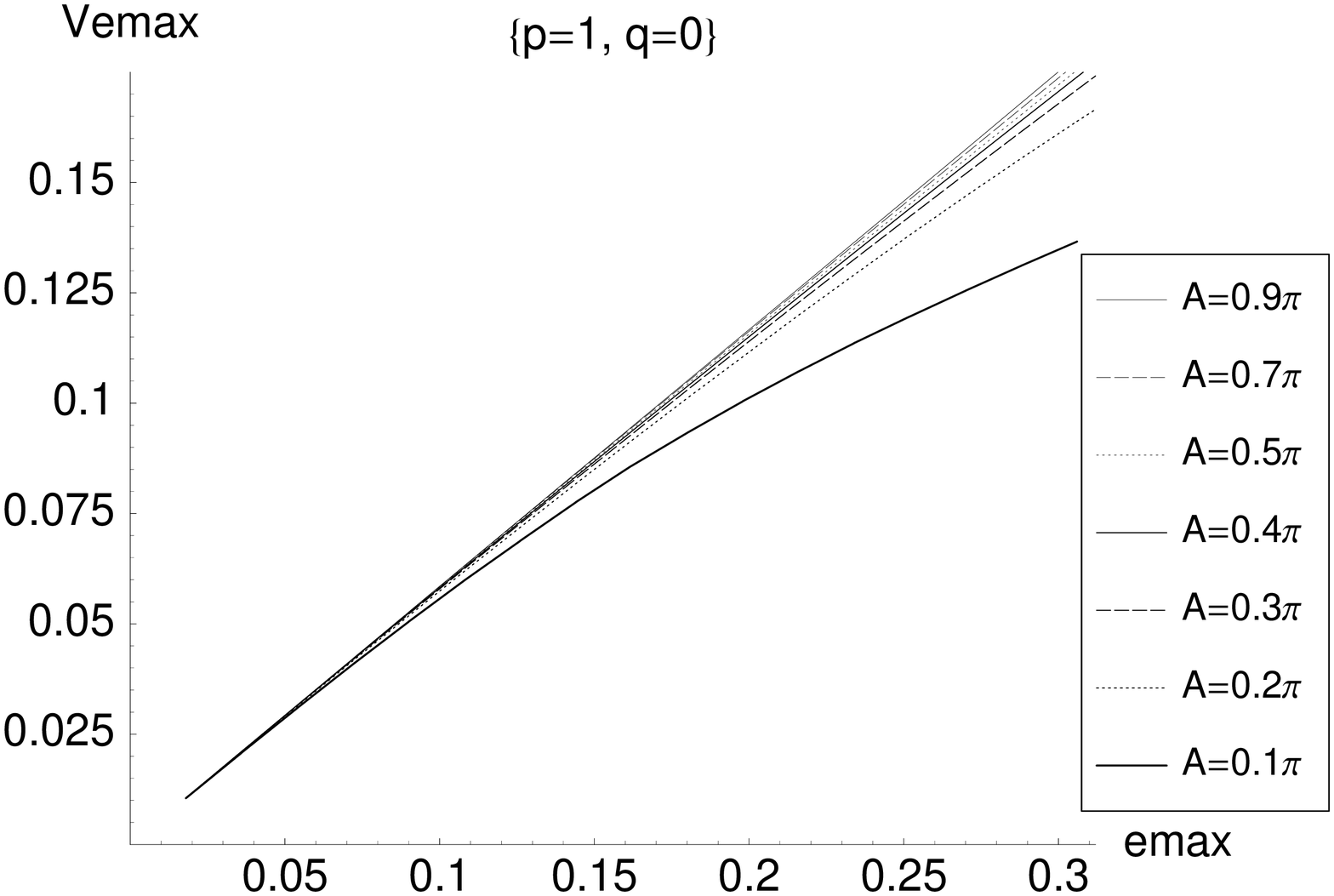} 
 \includegraphics[totalheight=0.3\textheight]{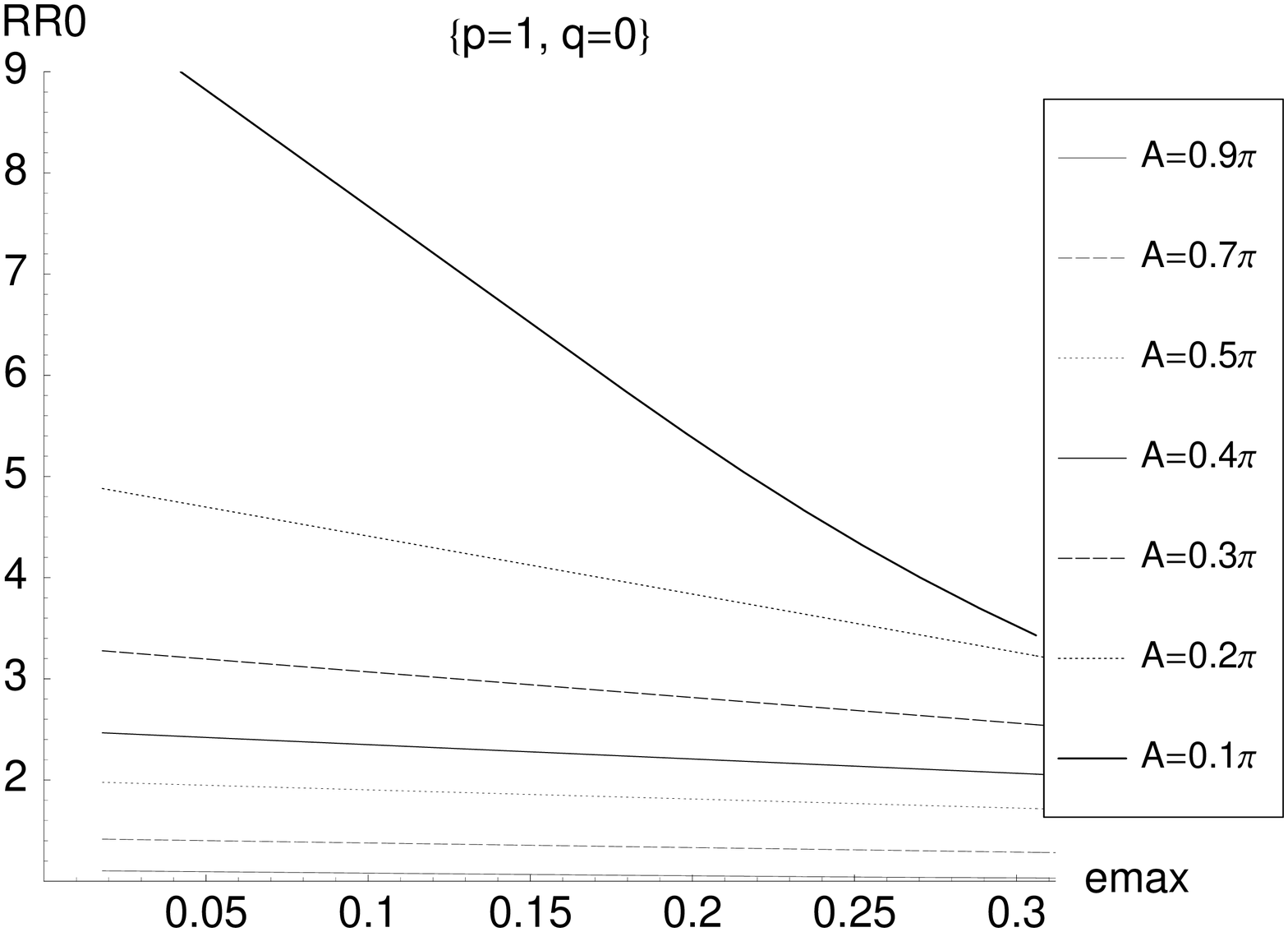} 
\end{center}
\caption{ 
 The case of Trojans.
 {\em Top:} collisional velocity $V_{imp}(e_{max})$ in units of $V_{kepler}$;
 cf. Fig.~\ref{fig_velemax}.
 {\em Bottom:} dimensionless collisional rate $\tilde{R}(e_{max})$;
 cf. Fig.~\ref{fig_rateemax}.
 Different curves correspond to different libration widths $A$.
 \label{fig_v_r_trojans}
 }  
\end{figure}

We now discuss a special case $q=0$, or 1:1 resonance, which corresponds to
a Trojan cloud of objects at one of the trigonal Lagrangian points.
All the formulas derived and discussed above are valid in this case, but the
results reveal important qualitative differences from the first- and higher-order resonances.
Figure~\ref{fig_v_r_trojans} presents the collisional velocity and the collisional rates
for Trojan clouds with different maximum eccentricities $e_{max}$
and different libration amplitudes $A$.

Like in the case $q>0$, $V_{imp}$ vanishes when $e_{max}$ goes to zero.
However, unlike for other resonances, $\tilde{R}(e_{max})$ has a maximum
at $e_{max} = 0$. What is more, that maximum collisional rate goes to infinity
when $A \rightarrow 0$.
Mathematically, this is explained in Appendix~\ref{Delta_integral_limit};
for instance, Eq.~(\ref{ultimate_limit}) is no longer valid.
Geometrically, the explanation is obvious (see Fig.~\ref{garden1}).
For $q>1$, even in the case where $e=0$ and $A=0$ the objects form an (infinitesimally
narrow) ring. For $q=0$, in that case the cloud simply shrinks to the
Lagrangian point~--- in other words, all the objects reside exactly at one and the same
point and have zero relative velocities.
Their volume density is infinitesimally high, and so is the collisional rate.
For small non-zero $e_{max}$ and $A$, frequent collisions are expected.
In contrast to this, 
the velocity would be lower, but the effect here is almost negligible,
because the velocities would be dominated by 3D-terms coming from inclinations,
which we do not consider in our model.
The implications for the Trojan asteroids in the solar system are discussed in the next Section.

%-----------------------------------------------------------------------
\section{Applications}
%-----------------------------------------------------------------------

The results may have a variety of astrophysical implications, whose detailed analysis
is beyond the scope of this paper.
Here we consider some of them and only briefly.

{\em 1. Resonant families of small bodies in the solar system}.
Our solar system is known to possess a number of small body populations locked in
mean-motion resonances. These are Trojan swarms of Jupiter and Trojans of other planets,
various asteroidal families in the main belt, Plutinos
and Twotinos in the Edgeworth-Kuiper belt.
We note again that the ``extreme'' analytics that we employed here required a number of
 simplifying assumptions.
For particular resonant families in the solar system, our model is 
obviously oversimplified to provide quantitatively accurate results.
Indeed, each of the families counts typically tens to hundreds of known objects, whose orbital 
elements are known at the individual level.
Therefore, to study their collisional evolution, direct $N$-body integration
\citep[e.g.][]{thebault-doressoundiram-2003}
or another version of statistical approach, developed by Dell'Oro 
and collaborators
\citep[e.g.][]{delloro-paolicchi-1998}, should be preferred.
Nevertheless, we make some quick comparisons of our results with those obtained
with more accurate methods \citep{marzari-et-al-1996,delloro-et-al-1998}.
In principle, we could choose any resonant group, as long as the number
of known members is sufficiently high to justify application of statistical approach.
We decide to consider Trojans, because for them,
as noted above, the effects of resonance on the collisional velocities 
and rates are the strongest.

Assuming for either of the two Jupiter Trojan swarms
$e_{max} = 0.1$
and $A =  0.1\pi \approx 20^\circ$,
Fig.~\ref{fig_v_r_trojans} suggests $\tilde{R} \approx 10$,
implying that the collisions are an order of magnitude more frequent than simple 
particle-in-a-box estimates would predict.
This result can be compared to the collisional rates among (non-resonant) main-belt asteroids.
If the latter had the same semimajor axes and the same distribution of inclinations
as Trojans, the rates would be 10 times lower.
According to Eq.~(\ref{R_0}), however ,
the collisional rate scales with the semimajor axis and mean inclination
(which enters the rate through $h$ or $\epsilon$ in Eq.~\ref{R_0}) as
$R \propto a^{-7/2} \langle i \rangle ^{-1}$.
The main-belt asteroids are roughly twice closer to the Sun than the Trojans,
and their mean inclinations of $\langle i \rangle = 8^\circ$
are twice smaller than that of Trojans ($15^\circ$).
This would give a factor of 20 lower rates, which would nearly compensate
the ten times higher $\tilde R$.
Therefore, the expected collisional rate for Trojans, despite their larger
heliocentric distance and broader vertical distribution, should be comparable with 
that for asteroids in the main belt.
Note that the result is rather sensitive to the choice of $e_{max}$ and $A$ and that
a more realistic distribution of eccentricities and resonant arguments can change the results
significantly.
With this caveat, our estimate is in a reasonable agreement with a result by 
\citet{marzari-et-al-1996} who found that the intrinsic collisional probabilities
for Trojans are about twice as high as in the main asteroidal belt.
Unlike the collision rate, the effect of the resonance on the collisional velocity is minor.
In fact, given rather large inclinations of Trojans, $V_{imp}$ will be dominated by 
inclination terms on the order $iV_{kepler}$ that are ignored in our treatment.
As a result, we expect nearly the same impact velocities for Trojans as for the main-belt asteroids,
and, as a consequence, the same collisional outcomes.
This fully agrees with conclusions of \citet{marzari-et-al-1996} and \citet{delloro-et-al-1998}
as well.

Similar estimates can, of course, be made for the
collisional evolution of Kuiper belt families.
For eccentricities lower than $e < 0.3$,
and the amplitude of libration of the resonant argument of tens of degrees,
the collisional rates may be up to a factor of two higher than for
non-resonant populations. The collisional velocities will not be
appreciably affected by the resonances.
Finally, applications to asteroidal families are also possible
\citep[cf.][]{marzari-et-al-1995,delloro-et-al-2001,delloro-et-al-2002}.
In this case, however, the theory developed in this paper needs to be adapted
to the case of internal resonances.

{\em 2. Clumpy debris disks.}
Our model may be particularly useful in the cases where orbital distributions
in the small body families are poorly known
or/and where the number of objects in the
family is far too large for $N$-body integrations to work.
The first fully applies to extrasolar ``asteroid belts'' around other
stars, whose existence is suggested by indirect, and very uncertain, methods.
The second applies directly to dust clouds such belts should produce.
This makes debris disks an ideal application for the model, as it enables
easy estimates for different central stars, parameters of presumed planets
and planetesimal belts, and various resonances in which planetesimals might be locked.
\citet{krivov-et-al-2006b} considered two possible scenarios
for formation of clumps observed in resolved debris disks.
In a standard scenario, the Poynting-Robertson force
delivers dust from outer regions of the disk to locations of external
mean-motion planetary resonances with a planet; dust grains get captured
and form visible clumps.
In another scenario \citep{wyatt-2003,wyatt-2006},
a population of invisible planetesimals resides in a resonance with
the planet, such as Plutinos and Trojans in the solar system;
the dust produced by these bodies would stay locked in the same resonance,
creating the dusty clumps.
With the help of simple analytic models,
\citet{krivov-et-al-2006b} showed that the first scenario works for disks
with the pole-on optical depths below about $\sim 10^{-4}$--$10^{-5}$.
Above this optical depth level, the first scenario will generate a narrow
resonant ring with a hardly visible azimuthal structure, rather than clumps.
The model of the first scenario was based on the particle-in-a-box formulas
for collisional rates, $\tilde{R}= 1$.  The fact that the corrections
to $\tilde{R}$ found here for the resonant case are only moderate lends
further support to the conclusions of that paper. On the other hand,
these corrections may lead to some quantitative changes. The ``critical''
optical depth, or fractional luminosity, of a debris disk in which
the standard scenario  may be efficient, shifts towards values that are by a factor
of several lower. 

%-----------------------------------------------------------------------
\section{Conclusions}
%-----------------------------------------------------------------------

The subject of this paper is  ``statistical'' celestial mechanics of the restricted
three-body problem. Specifically, we have investigated 
the averaged collisional velocities and rates of collisions
in an ensemble of planetesimals orbiting a star,
locked in a $(p+q):p$-resonance with a planetary 
perturber that revolves around the same primary, and having a distribution of eccentricities.

We use a statistical approach, based on the same philosophy as the one developed by
Dell'Oro and collaborators \citep{delloro-paolicchi-1998, delloro-et-al-1998}.
The main difference between the two approaches is that we treat a statistical
ensemble of pseudo-objects with continuous distributions of orbital elements,
rather than a finite set of objects with given orbital elements.

Our main findings are as follows:

(i) The mean collisional velocity is nearly the same
as in the case when the family is non-resonant.
In other words, it does not differ notably from the one predicted by the particle-in-a-box 
method: i.e., that 
velocity increases almost linearly with the maximum eccentricity $e_{max}$
of objects in the resonant family.

(ii) If the eccentricities of the objects are low to moderate, the mean collisional rates
are higher within a resonance than out of it. The enhancement in the collisional frequency
increases from shallow resonances to deeper ones, i.e.
with decreasing libration amplitude of the resonant argument.
For the libration widths of tens of degrees, the collisional rate may increase by about a factor of two.
There is a nonlinear dependence of the collisional rate
on $e_{max}$: the rate is highest at a certain $e_{max}$.
The smaller the resonant integer $p$ or the lower the order of resonance $q$, the smaller that $e_{max}$.
The maximum collisional rate is highest for the first-order resonances, $q = 1$.

(iii) In the special case of the primary, 1:1 resonance, the effect is the largest.
The collisional rates increase, and the velocities decrease, with decreasing eccentricities and/or
libration amplitudes in the family.

(iv) All the conclusions presented above apply to collisional velocities and rates
{\em averaged} over the swarm of objects. Therefore, the statement that rates and especially
velocities of collisions are only moderately affected by the resonant lock should not be
misinterpreted. Studies with $N$-body codes show that locally, namely at longitudes that correspond to 
the location of the clumps, both velocities and rates may be orders of magnitude higher
that outside the clumps \citep[see, e.g., Fig. 4 in][]{wyatt-2006}. Such studies show,
furthermore, that clumps are indeed regions where most of the collisions take place
and where most of the collisional debris emanate from, etc.
In contrast, our results are more pertinent to the global and long-term evolution of the whole 
resonant population, e.g. collisional lifetime of objects, mass outflow from the system
and so on. To illustrate the difference, let us discuss the lifetime of individual object.
It is not possible to say whether an individual planetesimal ``belongs to a clump'' or ``it does not''.
It spends part of its time in the clump and part out of it. Averaging over the orbits is
needed to predict its fate statistically, and that is exactly what we have done in this paper.

\bigskip
The results described above rest upon a number of simplifying assumptions that were made
to keep the problem analytically tractable. An incomplete list of them includes circular planetary 
orbit, treatment in the 2D, like-sized objects,
taking the 
same semimajor axis for all orbits, uniform and 
constant distribution of eccentricities and resonant arguments, a single isolated resonance without
interaction with background non-resonant objects and other resonant families.
However, even with all those assumptions, the mathematical complexity of the derivations is rather 
high, so that the prospects to relaxing most of them are questionable.
 From 
this viewpoint, we have demonstrated that
an analytic approach has severe limitations, both regard to the accuracy of results and
their applicability range.
Nevertheless, our model has a useful genericity, allowing one to quickly
estimate the strength of the resonant effects on the collisional evolution for various resonances, 
different libration amplitudes, different dispersions of eccentricities, etc.
Last but not least, as with any analytic study of a simplified problem, it offers
a clear dynamical explanation of why and how the resonant lock alters the rates
and velocities of collisions between the objects.

%------------------------------------------------------------------
% Acknowledgments
%------------------------------------------------------------------

\begin{acknowledgements}

We appreciate numerous stimulating discussions with Torsten L\"ohne
and useful conversations with Remy Reche,
Jean-Charles Augereau, Herv\'e Beust, and Mark Wyatt.
We thank Valerio Carruba and the anonymous referee
for their helpful and constructive reviews.
Martina Queck is funded by a graduate student fellowship of the Thuringia State.
Miodrag Srem{\v c}evi{\' c} is supported by the Cassini UVIS project.

\end{acknowledgements}

%------------------------------------------------------------------
% Appendix
%------------------------------------------------------------------
\newpage

\begin{appendix}

%-----------------------------------------------------------------------
\section{Transformation of $\Delta$-integral}
%-----------------------------------------------------------------------

Here, we will transform Eq.~(\ref{deltak_general}) to the form suitable for
efficient numerical calculations.

Two out of four integrations in Eq.~(\ref{deltak_general}), those over
$\overline{\omega}_2$ and $\theta_2$, can be done immediately with the
help of the two-branch collision condition.
Since $\cos\theta_2^+=\cos\theta_2^- = \cos\theta_2^c$ and
$|\sin\theta_2^+|=|\sin\theta_2^-| = |\sin\theta_2^c|$
we obtain:
\begin{eqnarray}
  &&\Delta^{(k)}(e_1,e_2)
  \nonumber\\
  &=&
  \frac{1}{a^2e_2(1-{{e_2}}^2)^2}
  \int_0^{2\pi}{\rm d}\theta_1
  \frac{(1+e_2\cos\theta_2^c)^3}{|\sin \theta_2^c|}
  \nonumber\\
  &\times&
  \sum_{\pm}
  {{V_{imp}^\pm}}^{k}(e_1,\theta_1,e_2)
  \int_0^{2\pi}{\rm d}\overline{\omega}_1
  \phi_{\overline{\omega}}(\overline{\omega}_1,\theta_1)
  \phi_{\overline{\omega}}(\overline{\omega}_2^c,\theta_2^c).
\end{eqnarray}
From now on, the superscript ``$c$'' of $\overline{\omega}_2^c$ and $\theta_2^c$
will be omitted. We stress, however, that both variables are functions
of $e_1$, $\overline{\omega}_1$ and $\theta_1$, and $e_2$, as calculated
from Eqs.~(\ref{omega_2^c}) and (\ref{theta_2^c}).
The $\Delta$-integral takes the form 
\begin{eqnarray}
  &&\Delta^{(k)}(e_1,e_2) 
  \nonumber\\
  &=&
  \frac{(1-e_1^2)^{1/2}(1-e_2^2)^{1/2}}{16\pi^2A^2a^2}
  \int_0^{2\pi}{\rm d}\theta_1
  \frac{1}{|e_2\sin \theta_2|(1+e_1\cos\theta_1)}
\nonumber\\
  &\times&
  \sum_{\pm}
  {{V_{imp}^\pm}}^{k}(e_1,\theta_1,e_2)
  \int_0^{2\pi}{\rm d}\overline{\omega}_1
  ~H_1~H_2
\label{deltak_interm}
\end{eqnarray}
where the Heaviside functions 
\be
H_{j} =  H\left[\Phi_0-A < \Phi_{j} <\Phi_0 +A\right], \qquad j=(1,2)
\ee
describe libration of resonant arguments $\Phi_1$ and $\Phi_2$ around $\Phi_0$ (e.g. $\pi$ or $0$).

The integral  $\int {\rm d}\overline{\omega}_1  H_1 H_2$
is analytically solvable.
Resonant arguments $\Phi_1$ and $\Phi_2$ can be written as
\be
  H_{j} = H\left[ -A < \{ p\overline{\omega}_1 + \chi_{j} \}_{2\pi}<A\right], \qquad j=(1,2),
\ee
where $\{.\}_{2\pi}$ denotes a modulo $2\,\pi$ operation
which returns values between $[-\pi,\pi]$.
E.g. $\{0\}_{2\pi}=0$, $\{\pi/2\}_{2\pi}=\pi/2$, 
$\{-\pi/2\}_{2\pi}=\{3\pi/2\}_{2\pi}=-\pi/2$, etc.
The arguments in the above equations are
\begin{eqnarray}
  \chi_1 &=& (p+q)~M(e_1,\theta_1)~-~\Phi_0,\\
  \chi_2 &=& (p+q)~M(e_2,\theta_2)~-~\Phi_0~-~p(\theta_2-\theta_1).
\end{eqnarray}
Here, $M=M(e,\theta)$ is the mean anomaly given by
Eqs.~(\ref{M_E})--(\ref{E_theta}).

Let us first consider the integral 
$\int_0^{2\pi}{\rm d}\overline{\omega}_1 H_1$.
Here, the offset $\chi_1$ 
is unimportant 
and it is easy to see that the integral evaluates to
\begin{equation}
  \int_0^{2\pi}{\rm d}\overline{\omega}_1 H_1
  =
  \int_0^{2\pi}{\rm d}\overline{\omega}_1 H\left[ -A < \{ p\overline{\omega}_1  \}_{2\pi}<A\right]
  = 2A.
\end{equation}

Similarly, $\int {\rm d}\overline{\omega}_1 H_1H_2 $
can be simplified to
\begin{eqnarray}
  \int_0^{2\pi}{\rm d}\overline{\omega}_1 H_1H_2
  &=&
  \int_0^{2\pi}{\rm d}\overline{\omega}_1~
  H\left[ -A < \{ p\overline{\omega}_1  \}_{2\pi}<A\right]
\nonumber\\
  &\times&
  H\left[ -A < \{ p\overline{\omega}_1 + \Delta\chi \}_{2\pi}<A\right],
\end{eqnarray}
while  the shift $\Delta\chi=\left\{\chi_2-\chi_1\right\}_{2\pi}$ is
\be
  \Delta\chi
  =
  \left\{
    (p+q)~\left[M(e_2,\theta_2)-M(e_1,\theta_1)\right] - p(\theta_2-\theta_1)
  \right\}_{2\pi} .
\ee
This can be further simplified by noting that we are ``overlapping''
intervals $[-A,A]$ and $[\Delta\chi-A,\Delta\chi+A]$ in
modulo $2\pi$ space.
Therefore
\begin{eqnarray}
  &&
  \int_0^{2\pi}{\rm d}\overline{\omega}_1 H_1H_2
  =
  2~A~
  \left(1 ~-~ \frac{|\Delta\chi|}{2A}\right)~
  H\left[1 ~-~ \frac{|\Delta\chi|}{2A}\right]
  \qquad
\nonumber\\
  &&
  1+\frac{|\Delta\chi|}{2A}\le \frac{\pi}{A}
  .
\label{int_om_ha_ha}
\end{eqnarray}
For $A \le \pi/2$ this branch is always taken.
Another branch is
\be
  \int_0^{2\pi}{\rm d}\overline{\omega}_1 H_1H_2
  =
  2~A~
  \left(2 ~-~ \frac{\pi}{A}\right)
  ,\qquad
  1+\frac{|\Delta\chi|}{2A}\ge \frac{\pi}{A}
  .
\ee
Note that $|\Delta\chi|\le\pi$ and $A\le\pi$.
In the limiting case $A=\pi$ we
obtain $\int {\rm d}\overline{\omega}_1  H_1 H_2=2A$.

The final result for the first branch (always taken for $A \le \pi/2$)
is:
\begin{eqnarray}
  \Delta^{(k)}(e_1,e_2) &=&
  \frac{(1-e_1^2)^{1/2}(1-e_2^2)^{1/2}}{8\pi^2Aa^2}
\nonumber\\
  &\times&
  \int_0^{2\pi}{\rm d}\theta_1
  \frac{1}{|e_2\sin \theta_2|(1+e_1\cos\theta_1)}
\nonumber\\
  &\times&
  \sum_{\pm}
  {{V_{imp}^\pm}}^{k}(e_1,\theta_1,e_2)
  \left(1 ~-~ \frac{|\Delta\chi^\pm|}{2A}\right)~
\nonumber\\
  &\times&
  H\left[1 ~-~ \frac{|\Delta\chi^\pm|}{2A}\right]
  ,
\label{deltak_final}
\end{eqnarray}
and similar to the second branch.
The only remaining integral is evaluated numerically 
using a Monte-Carlo method.

The above calculation assumes that the distribution of the resonant argument
within the libration band is uniform, Eq.~(\ref{phiom}).
If necessary, they can be repeated with a more realistic distribution.
It is sufficient to replace the Heaviside function in Eq.~(\ref{phiom}) by
another function $f( \{\Phi - \Phi_0\}_{2\pi} )$, for instance
a trigonometric one or a Gaussian, and re-do the calculation described here.

A more detailed analysis of the above equations allows one to establish
several useful properties of the $\Delta$-integrals.

First, it is symmetric with respect to its arguments.
Although it is not quite evident from \eqref{deltak_final}, 
one can use Eq.~(\ref{theta_2^c}) to change the
integration variable from $\theta_1$ to $\theta_2$.
The result will be identical to $\Delta(e_2,e_1)$, and therefore
$\Delta(e_2,e_1)=\Delta(e_1,e_2)$.
Of course, we have also tested the symmetry numerically.

Second, \eqref{deltak_final} readily shows that the simplified $\Delta$ integral
depends on the libration width $A$, but does not depend on the libration center $\Phi_0$.

Third, it is possible to calculate limiting cases of
Eq.~\eqref{deltak_final} and to get an idea of how large the effect of
resonance on collisional velocities and rates could be (Appendix
\ref{Delta_integral_limit}).

%-----------------------------------------------------------------------
\section{Limiting cases of $\Delta$-integral}
\label{Delta_integral_limit}
%-----------------------------------------------------------------------

Here we calculate the double limit
of the $\Delta^{(k=1)}(e_1, e_2)$-integral at $A \rightarrow 0$ and at $e_1 \rightarrow e_2$.
The first corresponds to an ``exact'' resonance.
The second ``emulates'' averaging over both eccentricities in
$\Delta(e_1,e_2)$.

We start with the limit $A \rightarrow 0$ of Eq.~(\ref{int_om_ha_ha}):
\be
  \lim_{A\rightarrow 0}
  \int_0^{2\pi}{\rm d}\overline{\omega}_1 H_1H_2
  =
  4 ~ A^2 ~\delta( \Delta \chi ),
\ee
which completely eliminates the $A$ dependence, as
Eq.~(\ref{deltak_interm}) contains $A^{-2}$.
Thus the $A \rightarrow 0$ limit of the $\Delta$-integral is finite.
Furthermore, the $\delta$-function immediately resolves the remaining
integral over $\theta_1$  to an explicit expression since
\be
  \delta( \Delta \chi(\theta_1) )
  =
  \sum_{i}
  \left|\frac{\partial \Delta \chi(\theta_1)}{\partial\theta_1}\right|^{-1}
  \delta(\theta_1 - \theta_1^{(i)}),
\label{delta chi}
\ee
where $\theta_1^{(i)}$ are the roots of the
equation $\Delta \chi(\theta_1^{(i)})=0$.
In practice, the calculation of Eq.~(\ref{delta chi})
is cumbersome, as it requires inversion of the Kepler equation.

We now consider the $e_2 \rightarrow e_1$ limit,
confining our analysis to the case $\Delta^{(k=1)}$.
After some algebraic manipulation, it can be shown that
\begin{eqnarray}
  \lim_{e_2 \rightarrow e_1} \frac{{{V_{imp}}}^+}{|e_2\sin {{\theta_2}}^+|}
    &=& 0,
\label{+}
\\
  \lim_{e_2 \rightarrow e_1} \frac{{{V_{imp}}}^-}{|e_2\sin {{\theta_2}}^-|}
    &=& \frac{2}{\sqrt{1-e_1^2}}~V_{kepler}.
\label{-}
\end{eqnarray}
Here, the ``$\pm$'' branches of the collision condition have slightly different
meaning than in \eqref{cc1}--\eqref{cc4}.
With the superscript ``$+$'' we denote the case
${{\theta_2}}^{c} \rightarrow \theta_1$,
and with ``$-$'' we denote ${{\theta_2}}^{c} \rightarrow -\theta_1$.
For the remaining nonzero ``$-$'' branch,
the condition $\Delta \chi(\theta_1^{(i)})=0$ is satisfied
at $\theta_1^{(1)} = 0$ and $\theta_1^{(2)}=\pi$.
The integral $\Delta^{(1)}$ is again finite, except at points
$e_1=e_2=e_\star$ when
\be
  \left.
    \frac{\partial \Delta \chi(\theta_1)}{\partial\theta_1}
  \right|_{\theta_1 = \theta_1^{(i)}}
      = 0.
\label{delta_chi_prime}
\ee
At points $e_1=e_2=e_\star$ Eq.~(\ref{delta chi}) implies that $\Delta^{(1)}$
is infinite, and those values
correspond to maxima in Fig.~\ref{contourfig_rate}.
Note that the maximum rates, $\tilde{R} (e_\star, e_\star)$, will rise to infinity
when $A \rightarrow 0$.
However, the $A \rightarrow 0$ limit of $\Delta^{(1)}(e_1, e_2)$
{\em integrated} over $e_1$ and $e_2$, will stay finite. As a result, 
the maxima in Fig.~\ref{fig_rateemax} will {\em not} be infinitely high
when $A \rightarrow 0$.

Equation (\ref{delta_chi_prime}) reduces to
\be
  \left.
    (p+q)\frac{\partial M(e_\star,\theta_1)}{\partial \theta_1}
  \right|_{\theta_1 = \theta_1^{(i)}}
  =p.
  \label{d_M1_over_t1}
\ee
The case with $\theta_1^{(2)}=\pi$ does not have a solution,
while for
$\theta_1^{(1)}=0$ the equation takes the form
\be
{\sqrt{1 + e_{\star}} \over (1 -e_{\star})^{3/2} }
  =
  { p + q \over p}.
\label{e_star_appendix}
\ee
The equation $\Delta \chi(\theta_1^{(i)})=0$ usually has more than
two roots 
(e.g. for $e_1=e_2=0$ those are $\theta_1^{(j)}=j\pi/q$, 
where $j$ is integer and $|j|\le q$).
In certain cases these additional roots can also simultaneously
satisfy Eq.~(\ref{delta_chi_prime}) leading to additional maxima
(see  Fig.~\ref{contourfig_rate} middle and right).

\begin{figure*}[tbh!]
\begin{center}
 \includegraphics[width=0.32\textwidth]{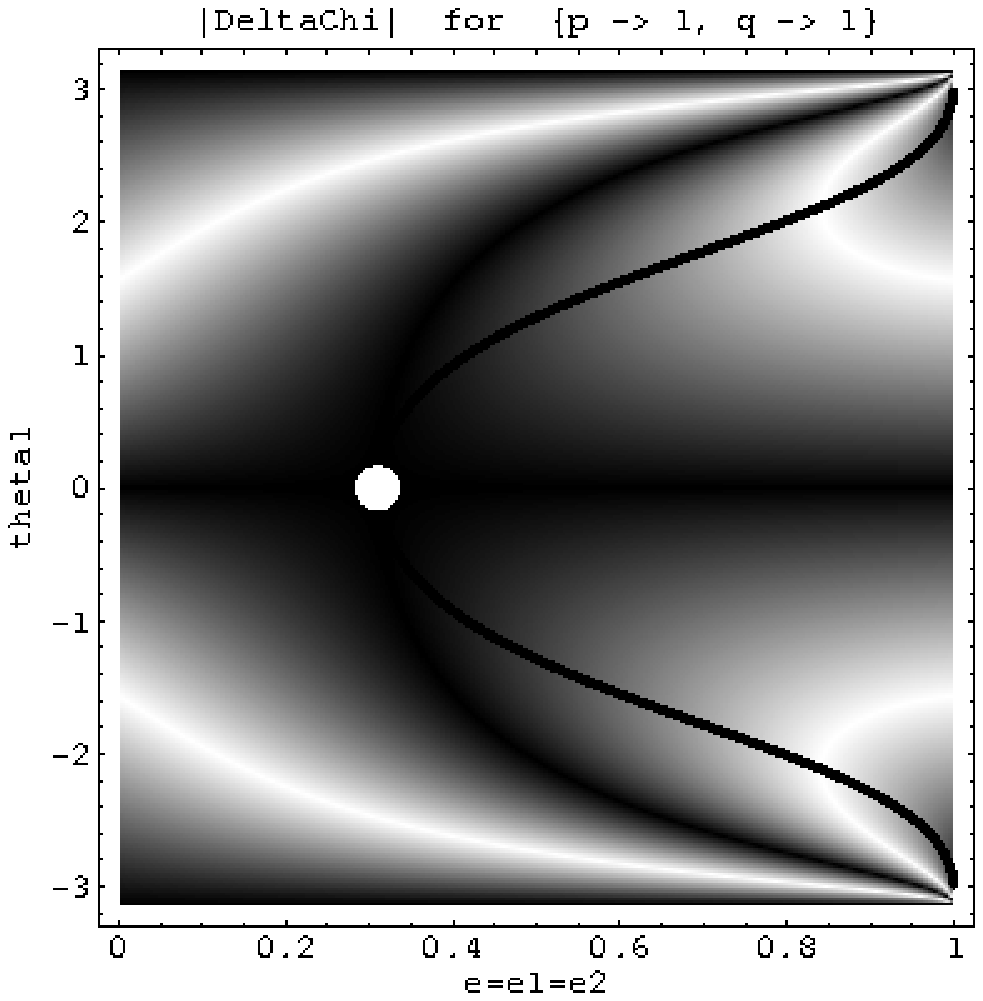}
 \includegraphics[width=0.32\textwidth]{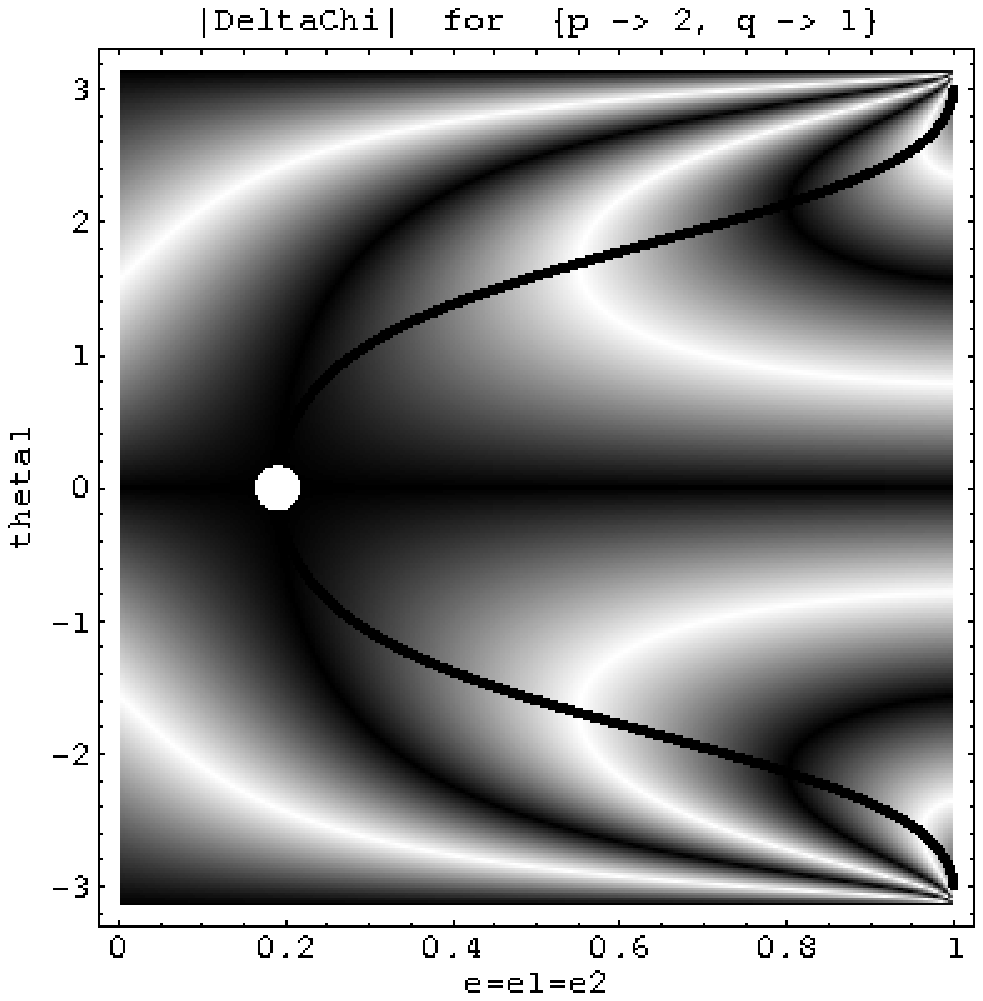}
 \includegraphics[width=0.32\textwidth]{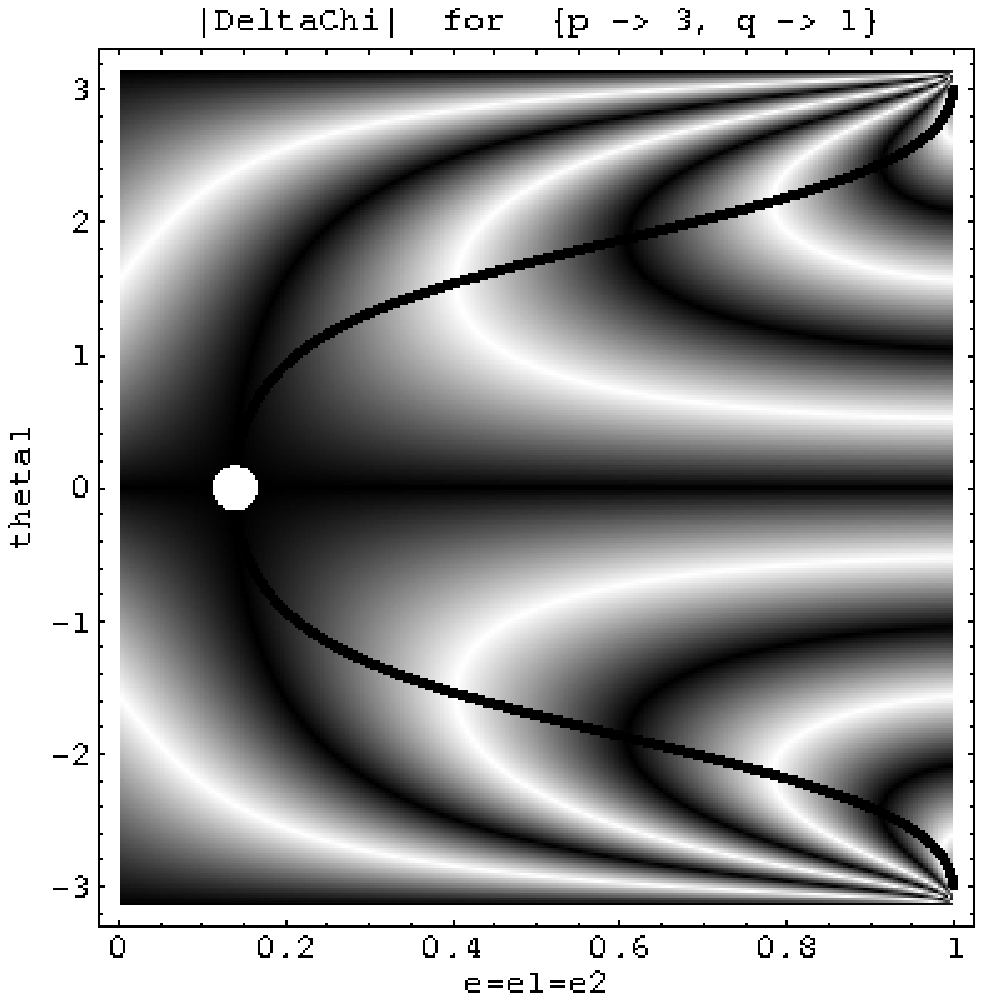} 
\end{center}
 \caption{The 2D plot of $|\Delta\chi(\theta_1,e_2=e_1)|$.
 From left to right: 2:1, 3:2, and 4:3 resonance.
 Black stripes in the density plot correspond to $\Delta\chi = 0$.
 The intersection between the black stripes and
 the solid line $(e_d, \theta_{1d})$ are the maxima.
 The white dot marks the first maximum, $(e_\star, 0)$.
 \label{fig_delta_chi}
 }
\end{figure*}

In order to find other maxima of the $\Delta$-integral,
it is easier to go back to Eq.~(\ref{d_M1_over_t1}).  
For $q\ge 1$ the only
solution to Eq.~(\ref{d_M1_over_t1}) is a pair
$(\theta_1=\theta_{1d},e_1=e_2=e_{d})$, where
\begin{eqnarray}
  x &=&
  \frac{(1-{{e_d}}^2)^{3/4}\sqrt{(p+q)/p}+e_d-1}{2e_{d}},\\
  \theta_{1d} &=&
  \pm 2 \arccos\left[\sqrt x\right].
  \label{CHI-theta1d}
\end{eqnarray}
It can be proved that $x>0$ and $dx/de_d<0$ (for $q/p\ge 0$ and
$0<e_d<1$), which has a consequence that $e_d=e_\star$ (when $x=1$ and
$\theta_{1d}=0$) is the smallest $e_d$ for which the $\Delta$-integral
has a maximum. Then, the other maxima (if any) can be obtained using
Eq.~(\ref{CHI-theta1d}) to check which pairs $(\theta_{1d},e_{d})$
satisfy $\Delta\chi =0$.
The density plot of $|\Delta\chi(\theta_1,e_2=e_1)|$, with the lines
$(\theta_1=\theta_{1d},e_1=e_2=e_{d})$ overplotted, is given in
Fig.~\ref{fig_delta_chi} for the same three resonances
as in Figs.~\ref{contourfig_vel} and \ref{contourfig_rate}: 2:1, 3:2 and 4:3.

We finally note that for $q \ge 1$,
\be
  \lim_{A \rightarrow 0, e_1 \rightarrow 0, e_2 \rightarrow 0}
  \Delta^{(1)} = \frac{V_{kepler}}{ 2 \pi^2 a^2} .
\label{ultimate_limit}
\ee
All of the equations in these Appendices hold for the case $q=0$
(1:1 resonance, Trojans) as well, except when an equation has a explicit $1/q$.
In particular, $q=0$ in Eq.~(\ref{e_star_appendix}) leads to $e_\star=0$,
which agrees with Fig.~\ref{fig_v_r_trojans}.

So far, we considered the  integral $\Delta^{(k=1)}$.
Of course, the integral $\Delta^{(k=0)}$ is not less important, because it
is needed to calculate the collisional velocity,
see. Eq.~(\ref{vimp_e1e2}).
For $k=0$ the "+" branch (Eq.~\ref{+}) diverges
when $e_1 \rightarrow e_2$.
Therefore,  $\Delta^{(k=0)}$ also diverges, and the collisional
velocity (\ref{vimp_e1e2}) goes to zero, as it should (see Fig.~\ref{fig_vele1}).

\end{appendix}

%------------------------------------------------------------------
% Bibliography
%------------------------------------------------------------------

\newcommand{\AAp}      {Astron. Astrophys.}
\newcommand{\AApSS}    {AApSS}
\newcommand{\AApT}     {Astron. Astrophys. Trans.}
\newcommand{\AdvSR}    {Adv. Space Res.}
\newcommand{\AJ}       {Astron. J.}
\newcommand{\AN}       {Astron. Nachr.}
\newcommand{\ApJ}      {Astrophys. J.}
\newcommand{\ApSS}     {Astrophys. Space Sci.}
\newcommand{\ARAA}     {Ann. Rev. Astron. Astrophys.}
\newcommand{\ARevEPS}  {Ann. Rev. Earth Planet. Sci.}
\newcommand{\BAAS}     {BAAS}
\newcommand{\CelMech}  {Celest. Mech. Dynam. Astron.}
\newcommand{\EMP}      {Earth, Moon and Planets}
\newcommand{\EPS}      {Earth, Planets and Space}
\newcommand{\GRL}      {Geophys. Res. Lett.}
\newcommand{\JGR}      {J. Geophys. Res.}
\newcommand{\MNRAS}    {Mon. Not. Roy. Astron. Soc.}
\newcommand{\PSS}      {Planet. Space Sci.}
\newcommand{\SSR}      {Space Sci. Rev.}

%--- For compilation without database

\end{document}